# Generalized Theory of Landau Damping

Boris V. Alexeev
Moscow Academy of Fine Chemical Technology (MITHT)
Prospekt Vernadskogo, 86, Moscow 119570, Russia
B.Alexeev@ru.net

Collisionless damping of electrical waves in plasma is investigated in the frame of the classical formulation of the problem. The new principle of regularization of the singular integral is used. The exact solution of the corresponding dispersion equation is obtained. The results of calculations lead to existence of discrete spectrum of frequencies and discrete spectrum of dispersion curves. Analytical results are in good coincidence with results of direct mathematical experiments.

## 1. Introduction

The collisionless damping of electron plasma waves was predicted by Landau in 1946 [1,2] and later was confirmed experimentally. Landau damping plays a significant role in many plasma experiments and belongs to the most well known phenomenon in statistical physics of ionized gases.

The physical origin of the collisionless Landau wave damping is simple. Really, if individual electron of mass $m_e$ moves in the periodic electric field, this electron can diminish its energy (electron velocity larger than phase velocity of wave) or receive additional energy from the wave (electron velocity less than phase velocity of wave). Then the total energy balance for a swarm of electrons depends on quantity of "cold" and "hot" electrons. For the Maxwellian distribution function, the quantity of "cold" electrons is more than quantity of "hot" electrons. This fact leads to, so-called, the collisionless Landau damping of the electric field perturbation.

In spite of transparent physical sense, the effect of Landau damping has continued to be of great interest to theorist as well. Much of this interest is connected with counterintuitive nature of result itself coupled with the rather abstruse mathematical nature of Landau's original derivation (including so-called Landau's rule of complex integral calculation). Moreover for these reason there were publications containing some controversy over the reality of the phenomenon (see for example, [3,4]).

In this paper I hope to clarify difficulties originated by Landau's derivative. The following consideration leads to another solutions of Landau equation, these ones in agreement with data of experiments. The problem Landau damping can be considered from the positions of Generalized Boltzmann Physical Kinetics [5] as the corresponding asymptotic solution of generalized Boltzmann equation. But here on purpose all consideration will be based on the classical Boltzmann equation written for collisionless case. The following results can be considered also as comments and prolongation of the materials published in the mentioned author`s monograph [5].

Let us remind the classical formulation of the problem. The usual derivation of Landau's damping begins by linearizing the Vlasov equation for the infinite homogeneous collisionless plasma. In doing so, we will make the same assumptions that were used in the Landau derivation namely:

(a) The integral collision term in Boltzmann equation (BE) is neglected;

(b) The evolution of electrons in a self-consistent electric field corresponds to a non-stationary one-dimensional model;

(c) Ions are in rest, the distribution functions (DF) for electrons $f_e$ deviates small from the maxwellian function $f_{0e}$;

$$f_e = f_{0e}(u) + \delta f_e(x,u,t), \tag{1.1}$$

(d) A wave number $k$ and complex frequency $\omega$ are appropriate to the wave mode considered;

$$\delta f_e = \langle \delta f_e \rangle e^{i(kx-\omega t)}, \tag{1.2}$$

(e) The quadratic terms determining the deviation from the equilibrium DF in kinetic equation

$$\frac{\partial f_e}{\partial t} + u\frac{\partial f_e}{\partial x} + F_e\frac{\partial f_e}{\partial u} = 0\,, \tag{1.3}$$

are neglected.

(f) The change of the electrical potential corresponds to the same spatial-time dependence as the perturbation of DF

$$\varphi = \langle \varphi \rangle e^{i(kx-\omega t)}, \tag{1.4}$$

(g) The self-consistent force $F_e$ is not too large

$$F_e = \frac{e}{m_e}\frac{\partial \varphi}{\partial x}, \tag{1.5}$$

where $e$ - absolute electron charge. It means that for calculation of function $F_e\left(\partial f_e / \partial u\right)$ the equilibrium DF is sufficient:

$$F_e\frac{\partial f_e}{\partial u} = \frac{e}{m_e}\frac{\partial \varphi}{\partial x}\frac{\partial f_{0e}}{\partial u} \tag{1.6}$$

Let us write down the complex frequency $\omega$ in the form

$$\omega = \omega' + i\omega'' \tag{1.7}$$

It means that

$$\delta f_e = \langle \delta f_e \rangle e^{i(kx-\omega' t)} e^{\omega'' t}. \tag{1.8}$$

As result the Landau's question can be formulated as follows – is it possible to find the solution of Eq. (1.3) by the all formulated restrictions, if $\omega'' < 0$? It is well known that the answer of this question leads to the necessity of solution of the following dispersion equation

$$\langle \delta n_e \rangle = \frac{e}{m_e}\langle \varphi \rangle k \int_{-\infty}^{+\infty} \frac{\partial f_{0e}/\partial u}{\omega - ku} du\,. \tag{1.9}$$

After using of maxwellian function for the one dimensional case and after introducing the dimensionless values

$$t = \frac{u\sqrt{m_e}}{\sqrt{2k_B T}},\ z_0 = \frac{\omega\sqrt{m_e}}{k\sqrt{2k_B T}}, \tag{1.10}$$

the dispersion equation (1.9) takes the standard form

$$\int_{-\infty}^{+\infty} \frac{e^{-t^2}}{z_0 - t} dt - \frac{\sqrt{\pi}}{z_0} = r_D^2 k^2 \frac{\sqrt{\pi}}{z_0}, \tag{1.11}$$

where

$$r_D = \sqrt{\frac{k_B T}{4\pi n_e e^2}} \tag{1.12}$$

$k$ is the wave number, $r_D$ is Debye radius. Let us introduce also the useful notations

$$z_0 \equiv x + iy \equiv \hat{\omega}' + \hat{\omega}'', \tag{1.13}$$

where

$$\hat{\omega}' = \frac{\omega' \sqrt{m_e}}{k\sqrt{2k_B T}}, \; \hat{\omega}'' = \frac{\omega'' \sqrt{m_e}}{k\sqrt{2k_B T}}. \tag{1.14}$$

As we see the solution of the dispersion equation (1.11) depends significantly on parameter $r_D^2 k^2$ or $r_D^2 (2\pi)^2 / \lambda^2$. In the so-called long wave approximation when $r_D^2 k^2 << 1$, Eq. (1.11) can be additionally simplified:

$$\int_{-\infty}^{+\infty} \frac{e^{-t^2}}{z_0 - t} dt = \frac{\sqrt{\pi}}{z_0}. \tag{1.15}$$

It seems that after all these restrictions and simplifications the quest of solution could not lead to troubles but it is very far from reality. The main problem consists in evaluation of the Landau integral

$$L(z_0) = \int_{-\infty}^{+\infty} \frac{e^{-t^2}}{z_0 - t} dt. \tag{1.16}$$

In the worst case this complex singular integral can be evaluated numerically, but numerical estimation of (1.16) leads to numerical solution of (1.11). It is reasonable, but un-convenient way because in many cases the problem of Landau damping is only a part of more complicated analytical problem. In the following I intend to estimate the accuracy of the Landau's original proposition for the $L(z_0)$ evaluation and therefore the accuracy of so-called Landau's rule, which can be found practically in all plasma physics textbooks.

## 2. Evaluation of Landau integral.

The separation of the real and imagine parts of the integral $L(z_0)$ leads to the following formulae:

$$L(z_0) = U + iV = \int_{-\infty}^{+\infty} \frac{e^{-t^2}}{z_0 - t} dt, \tag{2.1}$$

$$U = \operatorname{Re} L(z_0) = \int_{-\infty}^{+\infty} \frac{(x-t)e^{-t^2}}{(x-t)^2 + y^2} dt, \; V = \operatorname{Im} L(z_0) = -y \int_{-\infty}^{+\infty} \frac{(x-t)e^{-t^2}}{(x-t)^2 + y^2} dt. \tag{2.2}$$

Integral $L(z_0)$ can be evaluated by numerical way, the results of corresponding calculations for real part $Re\, L(z_0)$ and $Im\, L(z_0)$ are shown on Figures 1 and 2. As we see the integral surfaces have very complicated character.

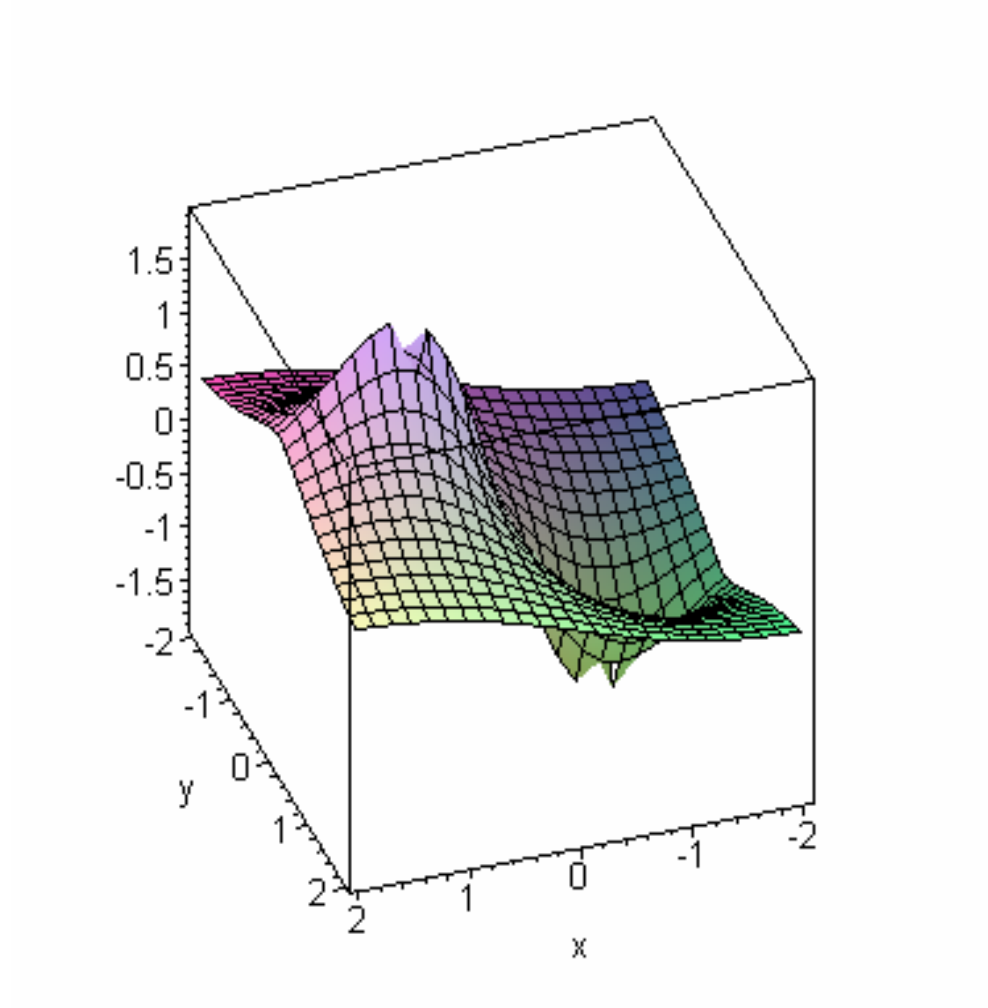

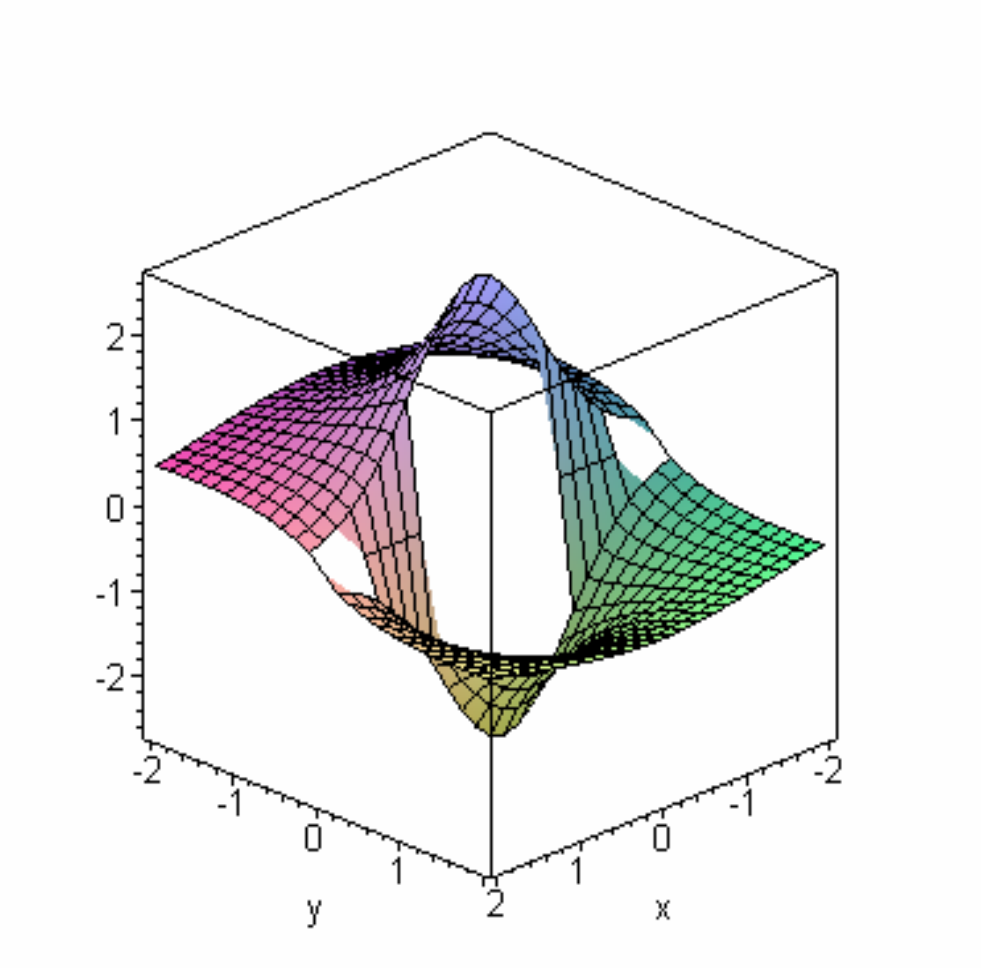

Fig. 1. $\operatorname{Re} L(z_0) = \int\limits_{-\infty}^{+\infty} \frac{(x-t)e^{-t^2}}{(x-t)^2 + y^2} dt$ in domain ( $x = -2,...,2; y = -2,...,2$ ), (on the left).

Fig. 2. $\operatorname{Im} L(z_0) = -y \int\limits_{-\infty}^{+\infty} \frac{(x-t)e^{-t^2}}{(x-t)^2 + y^2} dt$ in domain ( $x = -2,...,2; y = -2,...,2$ ), (on the right).

But the problem is not only in finding of numerical solution of (1.11).

*As it was mentioned above, the ideology of Landau damping penetrates in all physics (not only in plasma physics) and hundreds thousands of references connected with the consideration of this problem. This fact defines also the importance of analytical solution of the Landau problem.*

For the $L(z_0)$ evaluation the well known function (see for example the corresponding tables [6 - 8])

$$w(z_0) = e^{-z_0^2}\left(1 + \frac{2i}{\sqrt{\pi}} \int\limits_0^{z_0} e^{t^2} dt\right) \tag{2.3}$$

of the complex argument $z_0$ could be useful. Let us consider in details the connection of these functions $L(z_0)$ and $w(z_0)$. Introduce the substitution $t = z_0 + u$ in (2.1). The imaginary parts of $z$ and $u$ cancel each other because $t$ - real variable. But $z_0$ is fixed complex number $(z_0 = x + iy)$, then for complex variable $u$ the imagine part will be also constant and integration should be realized along a line $N$, which is parallel to the real axis:

$$L(z_0) = -e^{-z_0^2} \int\limits_N \frac{e^{-2z_0 u - u^2}}{u} du. \tag{2.4}$$

After introduction the function

$$f(z_0) = \int\limits_N \frac{e^{-2z_0 u - u^2}}{u} du \tag{2.5}$$

we have

$$\int\limits_{-\infty}^{+\infty} \frac{e^{-t^2}}{z_0 - t} dt = -e^{-z_0^2} f(z_0). \tag{2.6}$$

Differentiation (2.5) with respect to $z_0$ leads to result

$$f'(z_0) = -2\int_N e^{-2z_0 u - u^2} du \tag{2.7}$$

or, returning to the variable $t$:

$$f'(z_0) = -2\int_N e^{z_0^2 - t^2} dt\,. \tag{2.8}$$

But integral in the right-hand side of (2.8) contains Poisson integral, then

$$f'(z_0) = -2\sqrt{\pi} e^{z_0^2}\,. \tag{2.9}$$

Upon integrating (2.9) with respect to complex $z_0$:

$$f(z_0) = -2\sqrt{\pi}\int_0^{z_0} e^{s^2} ds + C\,, \tag{2.10}$$

where $C$ is constant, which should be defined. After comparison of (2.6) and (2.10), one obtains

$$\int_{-\infty}^{+\infty} \frac{e^{-t^2}}{z_0 - t} dt = e^{-z_0^2}\left(2\sqrt{\pi}\int_0^{z_0} e^{s^2} ds - C\right). \tag{2.11}$$

Let us find $C$. With this aim (2.11) is written by the conditions $x = 0,\ y \to +\infty$. In this case left-hand-side tends to zero and

$$C = 2\sqrt{\pi}\int_0^{iy} e^{s^2} ds\,,\ y \to +\infty\,. \tag{2.12}$$

Introduce the variable $v = is$. Because the integration is realized along imaginary axis, then $v$ is real value and

$$C = -2i\sqrt{\pi}\int_0^{-\infty} e^{-v^2} dv = \pi i, \tag{2.13}$$

After substitution of (2.13) in (2.11), one obtains for *upper half-plane* $(y > 0)$

$$\int_{-\infty}^{+\infty} \frac{e^{-t^2}}{z_0 - t} dt = e^{-z_0^2}\left(2\sqrt{\pi}\int_0^{z_0} e^{s^2} ds - \pi i\right) \tag{2.14}$$

or (for $y > 0$)

$$\int_{-\infty}^{+\infty} \frac{e^{-t^2}}{z_0 - t} dt = -\pi i w(z_0). \tag{2.15}$$

For *lower half-plane* the formula (2.11) should be transformed by another way. Let us return to the definition of constant $C$ and calculate this constant by the condition $x = 0,\ y \to -\infty$. In this case

$$C = 2\sqrt{\pi}\int_0^{iy} e^{s^2} ds,\ \text{by}\ y \to -\infty\,; \tag{2.16}$$

$$C = -2i\sqrt{\pi}\int_0^{+\infty} e^{-v^2} dv = -\pi i\,. \tag{2.17}$$

As result for *lower half-plane* $(y < 0)$

$$\int_{-\infty}^{+\infty} \frac{e^{-t^2}}{z_0 - t} dt = e^{-z_0^2} \left( 2\sqrt{\pi} \int_0^{z_0} e^{s^2} ds + \pi i \right). \quad (2.18)$$

It means that for the case of Landau damping when we try to find solution of Eq. (1.11) in *lower half-plane* ( $y < 0$ or $\hat{\omega}'' < 0$ ) Landau integral $L(z_0)$ cannot be written via function $w(z_0)$.

The comparison of (2.14) and (2.18) leads to conclusion that Landau integral function $L(z_0)$ has jump discontinuity in the vicinity of real axis. Let us consider this problem in details using “the approximation of small $y$ ”. Let us return to evaluation of the function $L(z_0)$ near the real axis with the help of (2.14). In the first approximation of the small $y$ $(z_0 = x + iy,\ 0 < y << 1)$ we have for $L(z_0) = U(z_0) + iV(z_0)$

$$U = -2\pi xye^{-x^2} + 2\sqrt{\pi} e^{-x^2} \int_0^x e^{s^2} ds, \quad (2.19)$$

$$V = -\pi e^{-x^2} + 2\sqrt{\pi} y - 4\sqrt{\pi} xye^{-x^2} \int_0^x e^{s^2} ds. \quad (2.20)$$

For the case $y < 0$, $|y| << 1$ from (2.18) follows

$$U = 2\pi xye^{-x^2} + 2\sqrt{\pi} e^{-x^2} \int_0^x e^{s^2} ds, \quad (2.21)$$

$$V = \pi e^{-x^2} + 2\sqrt{\pi} y - 4\sqrt{\pi} xye^{-x^2} \int_0^x e^{s^2} ds. \quad (2.22)$$

From relations (2.19) - (2.22) follow: A) Function $U$ by crossing real axis from the domain $y < 0$ in the domain where $y > 0$ has no jump discontinuity in the vicinity of real axis. B) Function $V$ by crossing real axis from the domain $y < 0$ in the domain where $y > 0$ has jump discontinuity in the vicinity of real axis, namely $V(x, y = -0) = \pi e^{-x^2}$, $V(x, y = +0) = -\pi e^{-x^2}$. In particular, $V(x = 0, y = -0) = \pi$, $V(x = 0, y = +0) = -\pi$.

As confirmation of validity of analytical asymptotic calculations Figures 3 – 8 demonstrate the practical identical results of calculations with the help of exact formulae (2.2) and its asymptotic variants (2.19) – (2.22).

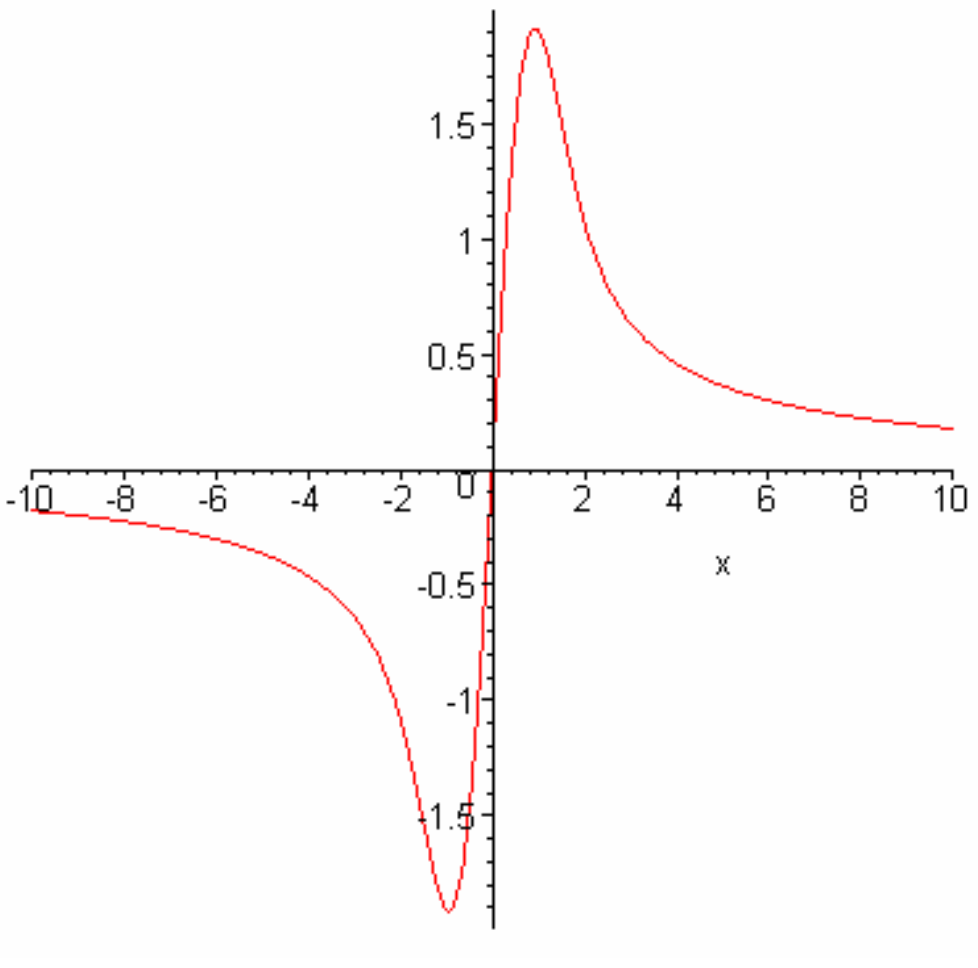

Fig. 3. Results of calculation $\mathrm{Re}\, L(z_0) = \int_{-\infty}^{+\infty} \frac{(x - t)e^{-t^2}}{(x - t)^2 + y^2} dt$ by $y = 0.001$.

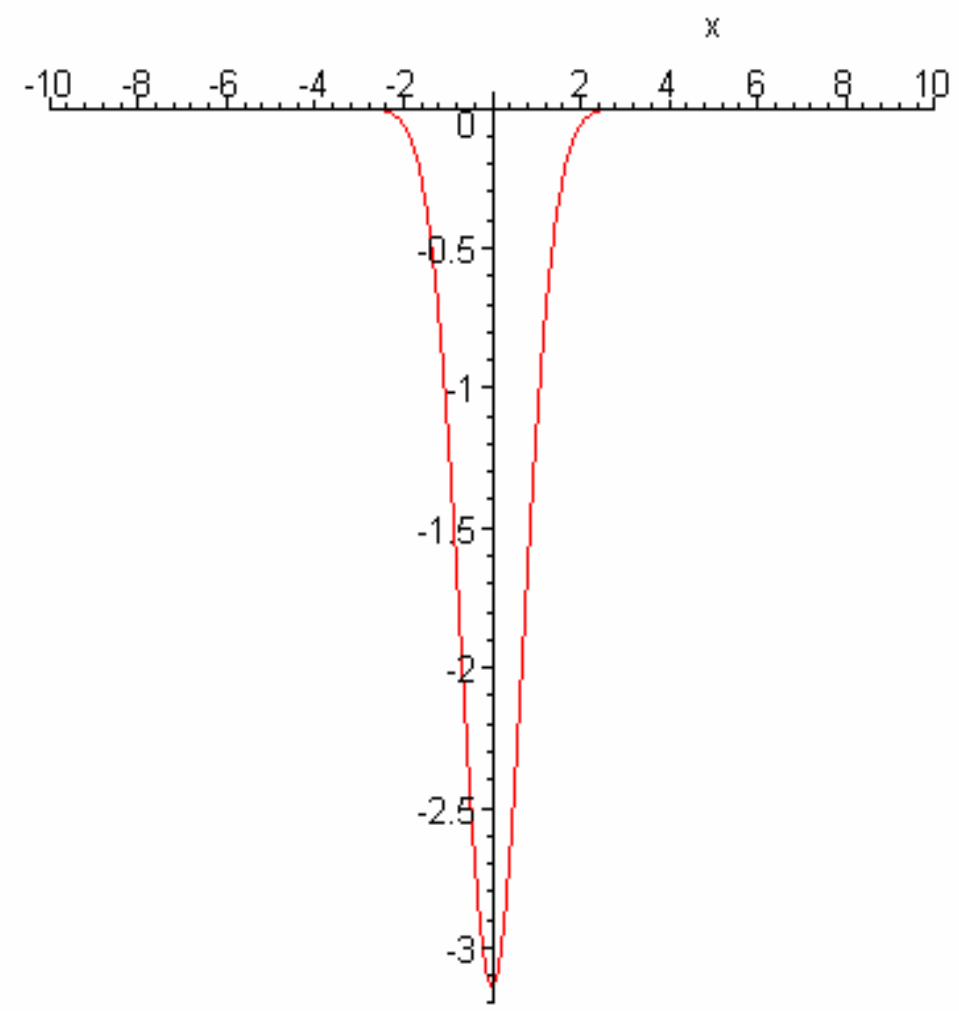

Fig. 4. Results of calculation $\operatorname{Im} L(z_0) = -y \int_{-\infty}^{+\infty} \frac{(x-t)e^{-t^2}}{(x-t)^2 + y^2} dt$ by $y = 0.001$.

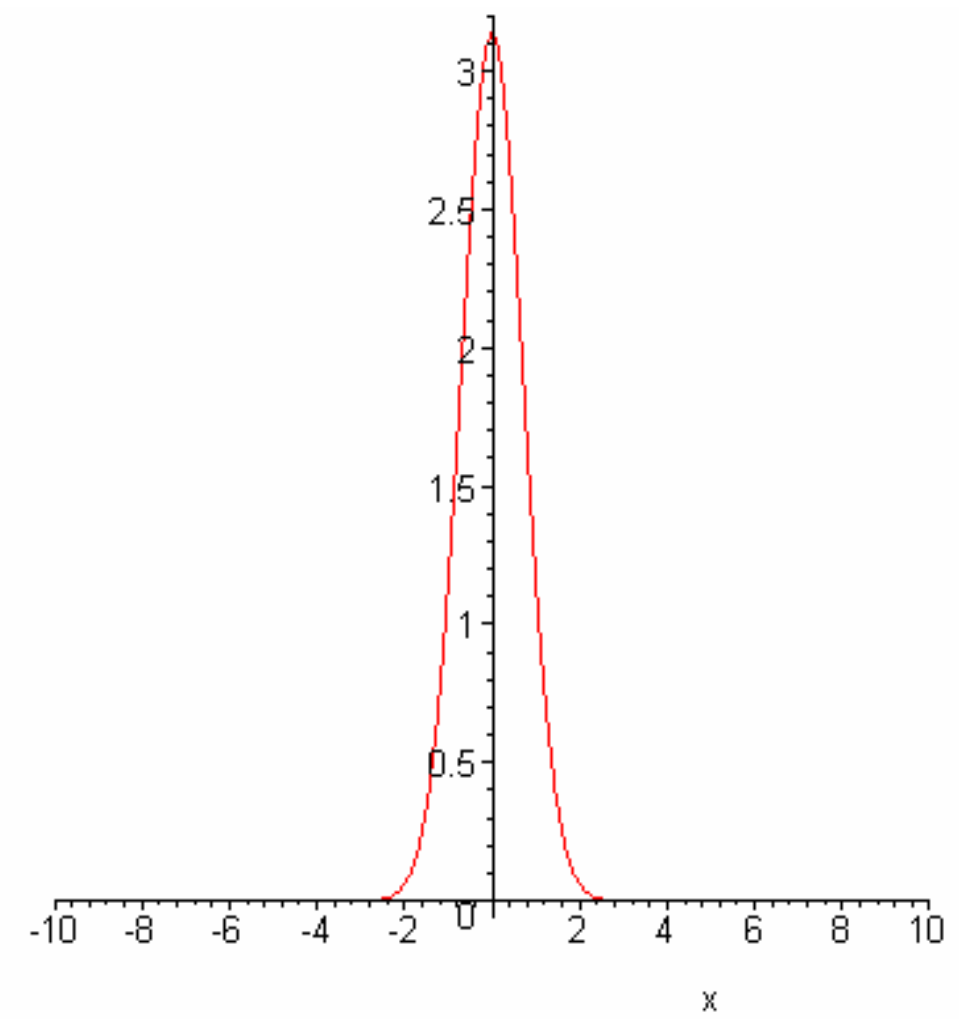

Fig. 5. Results of calculation $\operatorname{Im} L(z_0) = -y \int_{-\infty}^{+\infty} \frac{(x-t)e^{-t^2}}{(x-t)^2 + y^2} dt$ by $y = -0.001$.

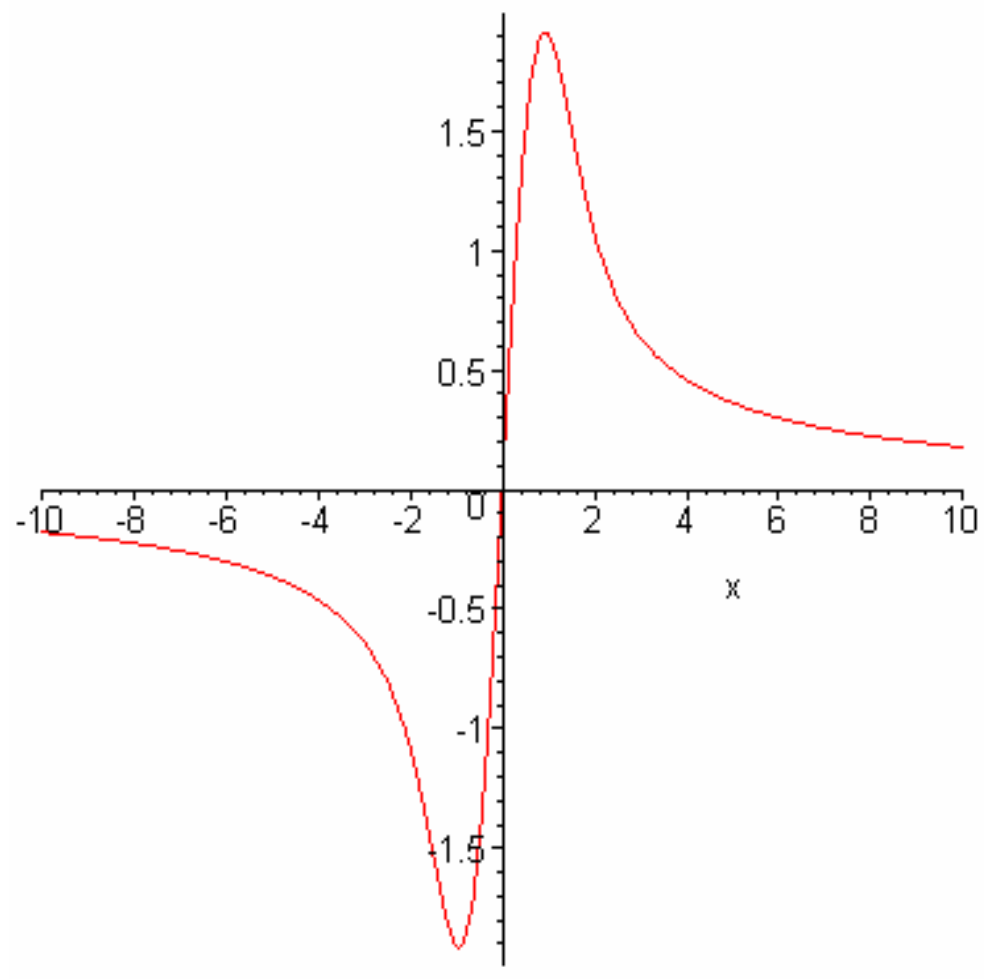

Fig. 6. Results of calculation $U(x, y = 0{,}001) = -2\pi x y e^{-x^2} + 2\sqrt{\pi} e^{-x^2} \int_0^x e^{s^2} ds$.

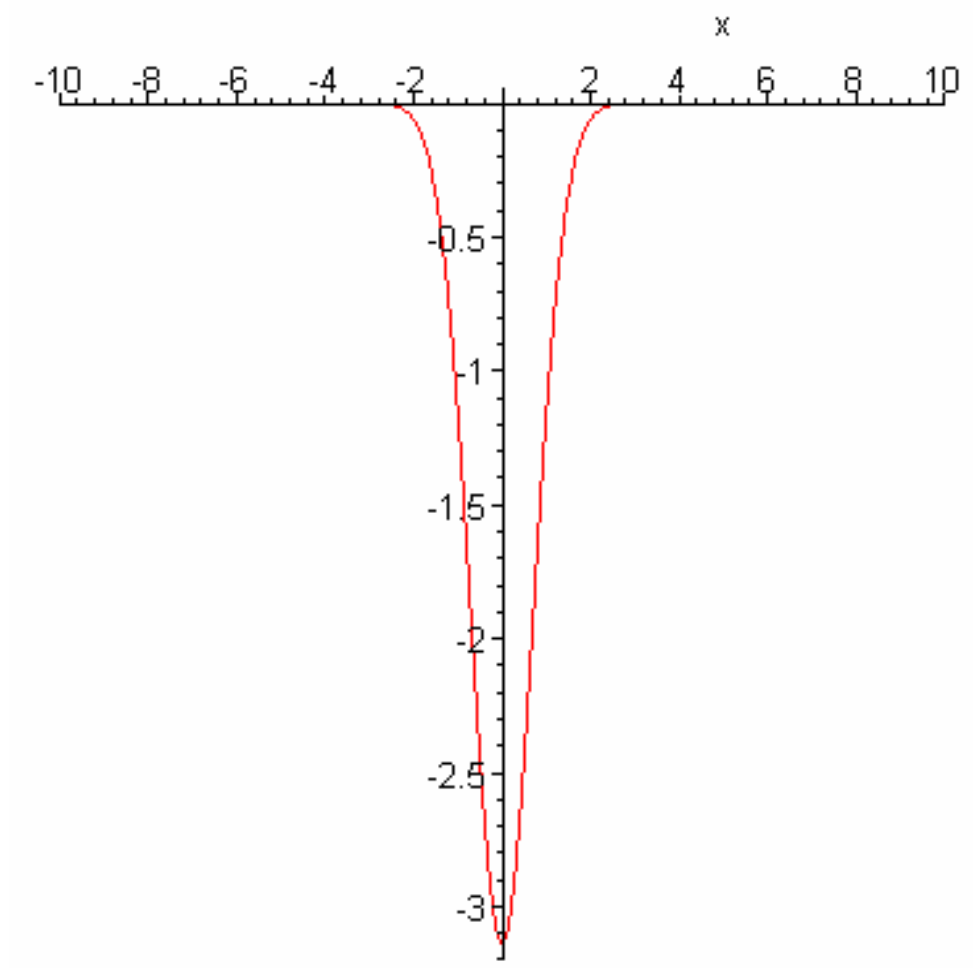

Fig. 7. Results of calculation $V(x, y = 0{,}001) = -\pi e^{-x^2} + 2\sqrt{\pi}\, y - 4\sqrt{\pi}\, xye^{-x^2} \int\limits_0^x e^{s^2} ds$.

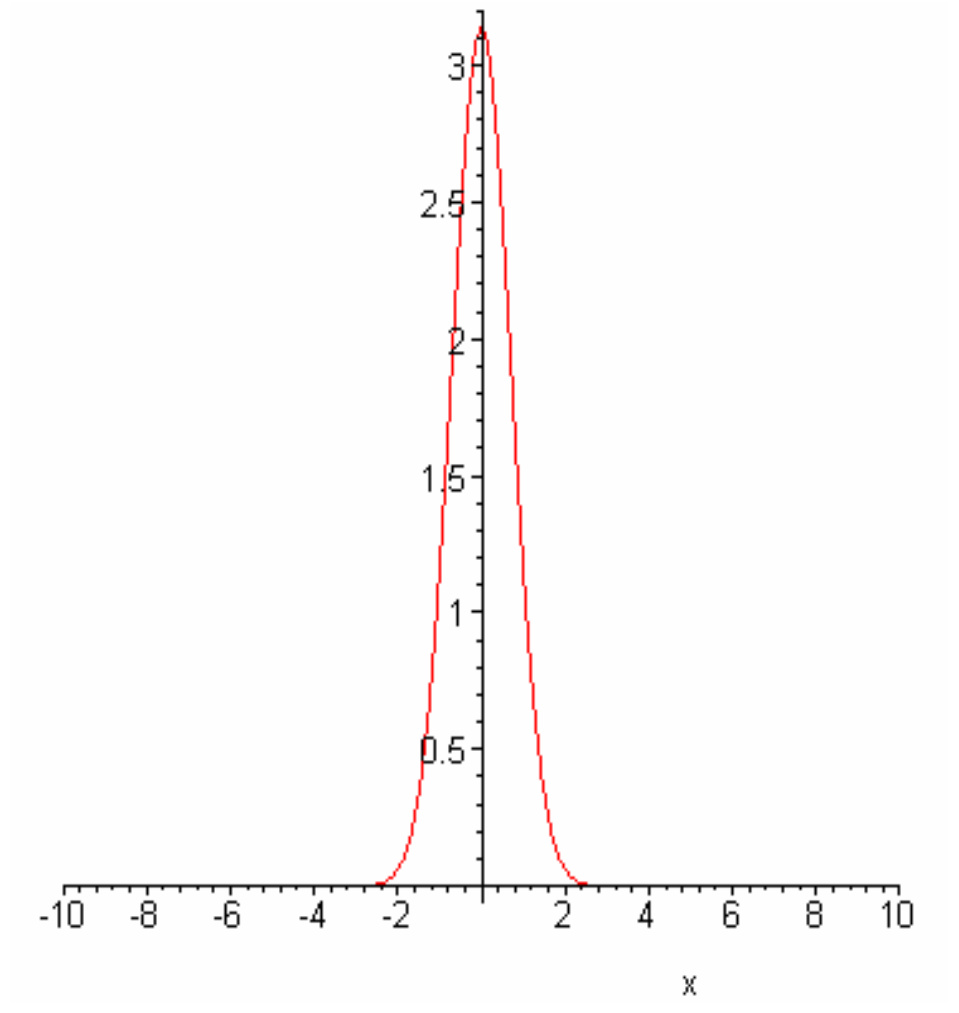

Fig. 8. Results of calculation $V(x, y = -0{,}001) = \pi e^{-x^2} + 2\sqrt{\pi}\, y - 4\sqrt{\pi}\, xye^{-x^2} \int\limits_0^x e^{s^2} ds$.

Let us consider now another possible approximation of the Landau integral $L(z_0) = \int\limits_{-\infty}^{+\infty} \frac{e^{-t^2}}{z_0 - t} dt$, namely "the approximation of large $z_0$". The formal expansion of function $\left(1 - \frac{t}{z_0}\right)^{-1}$ in complex series can be written as

$$\left(1 - \frac{t}{z_0}\right)^{-1} = 1 - \frac{t}{z_0} + \left(\frac{t}{z_0}\right)^2 - \left(\frac{t}{z_0}\right)^3 + \left(\frac{t}{z_0}\right)^4 - \ldots . \quad (2.23)$$

After substitution of (2.23) in Landau integral one obtains

$$L(z_0) = \int\limits_{-\infty}^{+\infty} \frac{e^{-t^2}}{z_0 - t} dt = \frac{1}{z_0} \sum_{n=0}^{\infty} \int\limits_{-\infty}^{+\infty} e^{-t^2} (-1)^n \left(\frac{t}{z_0}\right)^n dt . \quad (2.24)$$

But integrals in the right-hand-side of (2.24) containing odd powers of $\left(\frac{t}{z_0}\right)^{2n+1}$ turning into zero then

$$L(z_0)=\int_{-\infty}^{+\infty}\frac{e^{-t^2}}{z_0-t}dt=\frac{1}{z_0}\int_{-\infty}^{+\infty}e^{-t^2}\left(1+\left(\frac{t}{z_0}\right)^2+\left(\frac{t}{z_0}\right)^4+\ldots\right)dt\,. \qquad (2.25)$$

For convenience of the series convergence, write down the integral series (2.25) as

$$\int_{-\infty}^{+\infty}\frac{e^{-t^2}}{z_0-t}dt=\sum_{k=0}^{\infty}S_{2k+1}\,. \qquad (2.26)$$

In this case the subscript in the sum corresponds to maximum reversed power of $z_0$ in partial sum of infinite series. The first terms of the series have the form

$$S_1=\frac{\sqrt{\pi}}{z_0}, \qquad S_3=\frac{\sqrt{\pi}}{z_0}\left(1+\frac{1}{2}\frac{1}{z_0^2}\right), \qquad S_5=\frac{\sqrt{\pi}}{z_0}\left(1+\frac{1}{2}\frac{1}{z_0^2}+\frac{3}{4}\frac{1}{z_0^4}\right),$$

$$S_7=\frac{\sqrt{\pi}}{z_0}\left(1+\frac{1}{2}\frac{1}{z_0^2}+\frac{3}{4}\frac{1}{z_0^4}+\frac{15}{8}\frac{1}{z_0^6}\right),$$

$$S_9=\frac{\sqrt{\pi}}{z_0}\left(1+\frac{1}{2}\frac{1}{z_0^2}+\frac{3}{4}\frac{1}{z_0^4}+\frac{15}{8}\frac{1}{z_0^6}+\frac{105}{16}\frac{1}{z_0^8}\right),$$

$$S_{11}=\frac{\sqrt{\pi}}{z_0}\left(1+\frac{1}{2}\frac{1}{z_0^2}+\frac{3}{4}\frac{1}{z_0^4}+\frac{15}{8}\frac{1}{z_0^6}+\frac{105}{16}\frac{1}{z_0^8}+\frac{945}{32}\frac{1}{z_0^{10}}\right),$$

$$S_{13}=\frac{\sqrt{\pi}}{z_0}\left(1+\frac{1}{2}\frac{1}{z_0^2}+\frac{3}{4}\frac{1}{z_0^4}+\frac{15}{8}\frac{1}{z_0^6}+\frac{105}{16}\frac{1}{z_0^8}+\frac{945}{32}\frac{1}{z_0^{10}}+\frac{10395}{64}\frac{1}{z_0^{12}}\right),$$

$$S_{15}=\frac{\sqrt{\pi}}{z_0}\left(1+\frac{1}{2}\frac{1}{z_0^2}+\frac{3}{4}\frac{1}{z_0^4}+\frac{15}{8}\frac{1}{z_0^6}+\frac{105}{16}\frac{1}{z_0^8}+\frac{945}{32}\frac{1}{z_0^{10}}+\frac{10395}{64}\frac{1}{z_0^{12}}+\frac{135135}{128}\frac{1}{z_0^{14}}\right),$$

$$S_{17}=\frac{\sqrt{\pi}}{z_0}\left(1+\frac{1}{2}\frac{1}{z_0^2}+\frac{3}{4}\frac{1}{z_0^4}+\frac{15}{8}\frac{1}{z_0^6}+\frac{105}{16}\frac{1}{z_0^8}+\frac{945}{32}\frac{1}{z_0^{10}}+\frac{10395}{64}\frac{1}{z_0^{12}}+\frac{135135}{128}\frac{1}{z_0^{14}}+\frac{135135\cdot 15}{256}\frac{1}{z_0^{16}}\right)$$

$$S_{19}=\frac{\sqrt{\pi}}{z_0}\left(\begin{array}{l}1+\frac{1}{2}\frac{1}{z_0^2}+\frac{3}{4}\frac{1}{z_0^4}+\frac{15}{8}\frac{1}{z_0^6}+\frac{105}{16}\frac{1}{z_0^8}+\frac{945}{32}\frac{1}{z_0^{10}}+\frac{10395}{64}\frac{1}{z_0^{12}}+\frac{135135}{128}\frac{1}{z_0^{14}}+\frac{135135\cdot 15}{256}\frac{1}{z_0^{16}}\\+\frac{135135\cdot 15\cdot 17}{512}\frac{1}{z_0^{18}}\end{array}\right)$$

Consider the peculiar features of convergence of the series (2.26). Figures 9 – 17 contain the results of the $L(z_0)$ calculation for three successive approximations in the method of "large $z_0$" for real and imagine parts of the corresponding complex functions and sectional views for line $x=y$. As we see the convergence of successive approximations exist in domain $x,y\geq 3$.

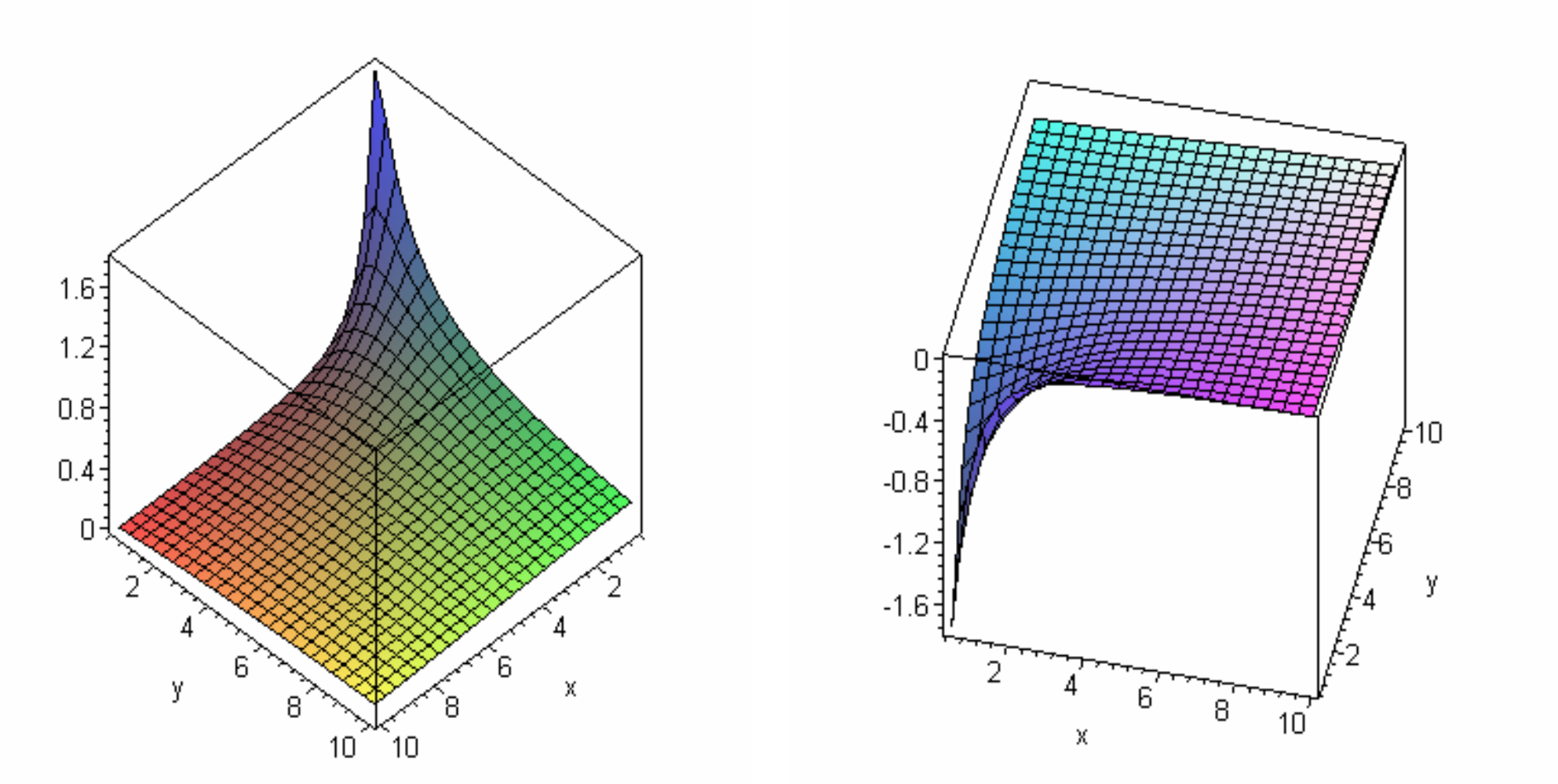

**Fig. 9. The 3D image of function** $\mathrm{Re}\dfrac{\sqrt{\pi}}{z_0} = \dfrac{x\sqrt{\pi}}{x^2+y^2}$**,** $(x = 0.5,\ldots,10;\ y = 0.5,\ldots,10)$ **(left) and function** $\mathrm{Im}\dfrac{\sqrt{\pi}}{z_0} = -\dfrac{y\sqrt{\pi}}{x^2+y^2}$**,** $(x = 0.5,\ldots,10;\ y = 0.5,\ldots,10)$ **(right).**

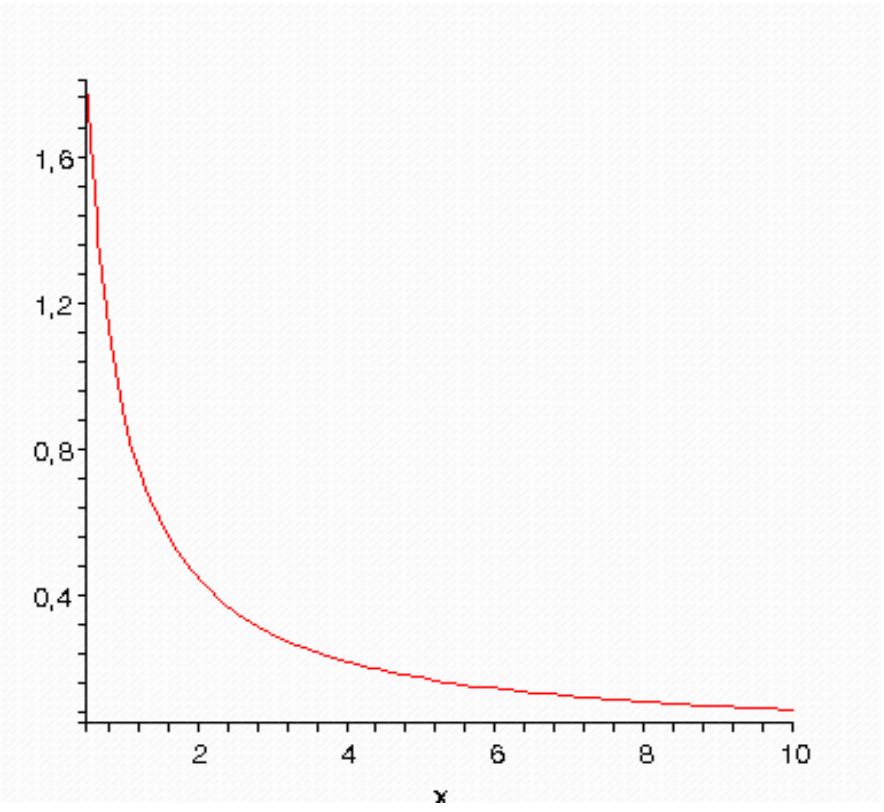

**Fig. 10. Sectional view of function** $\mathrm{Re}\dfrac{\sqrt{\pi}}{z_0} = \dfrac{x\sqrt{\pi}}{x^2+y^2}$**,** $(x = 0.5,\ldots,10;\ y = 0.5,\ldots,10)$ **for line** $x = y$**;**

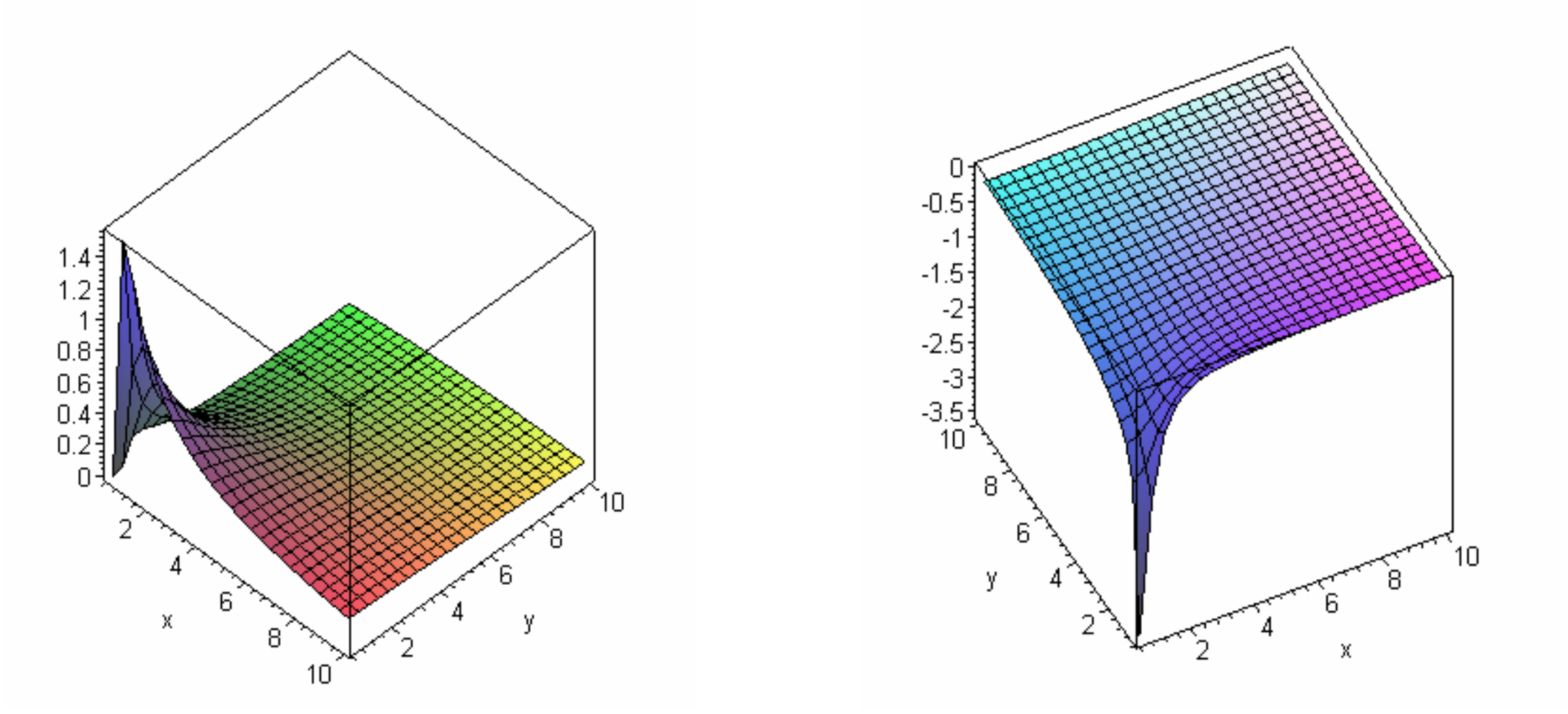

**Fig. 11. The 3D image of function** $\mathrm{Re}\frac{\sqrt{\pi}}{z_0}\left(1+0.5\frac{1}{z_0^2}\right)$, $(x=0.5,\ldots,10;\ y=0.5,\ldots,10)$ **(left).**

**Fig. 12. The 3D image of function** $\mathrm{Im}\frac{\sqrt{\pi}}{z_0}\left(1+0.5\frac{1}{z_0^2}\right)$, $(x=0.5,\ldots,10;\ y=0.5,\ldots,10)$ (right).

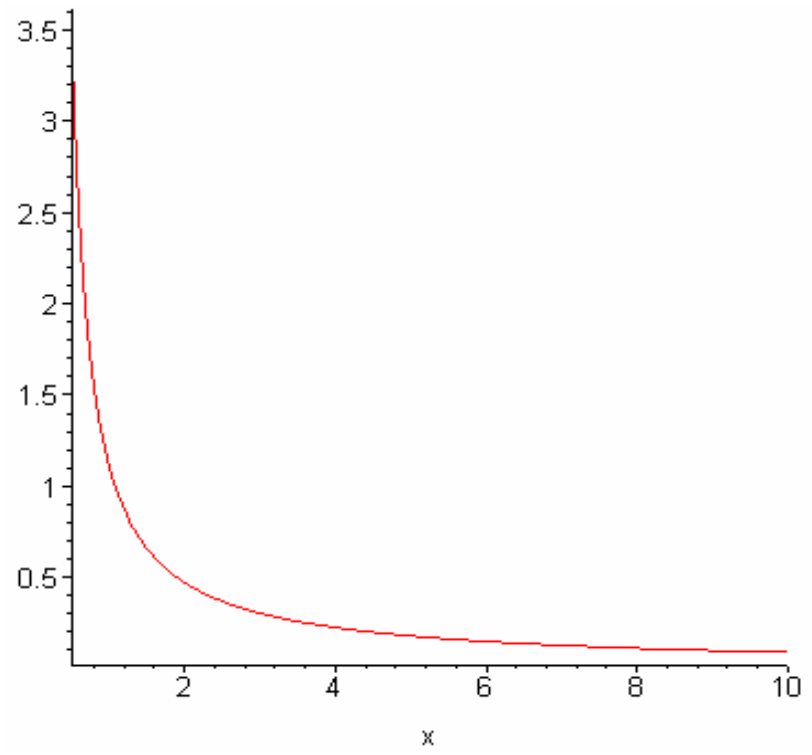

**Fig. 13. Sectional view of function** $\left|\mathrm{Im}\frac{\sqrt{\pi}}{z_0}\left(1+0.5\frac{1}{z_0^2}\right)\right|$ $(x=0.5,\ldots,10;\ y=0.5,\ldots,10)$ **for line** $x=y$ **.**

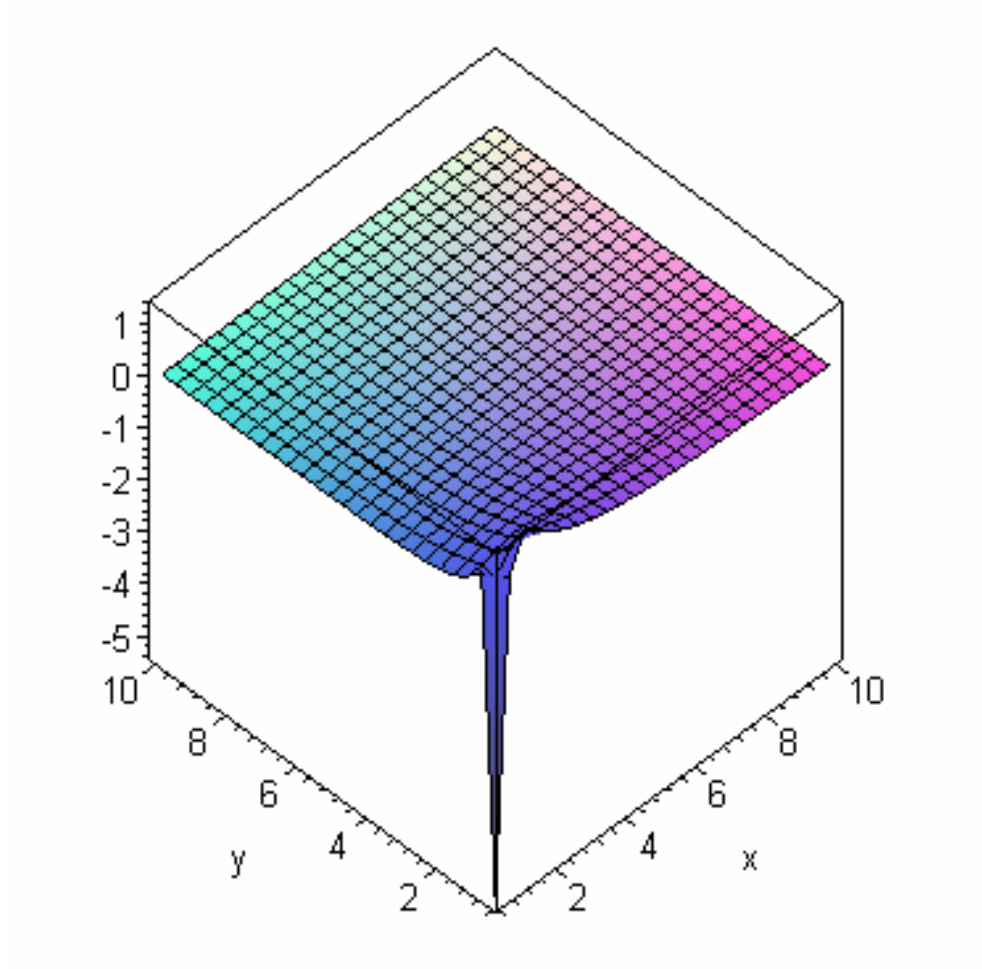

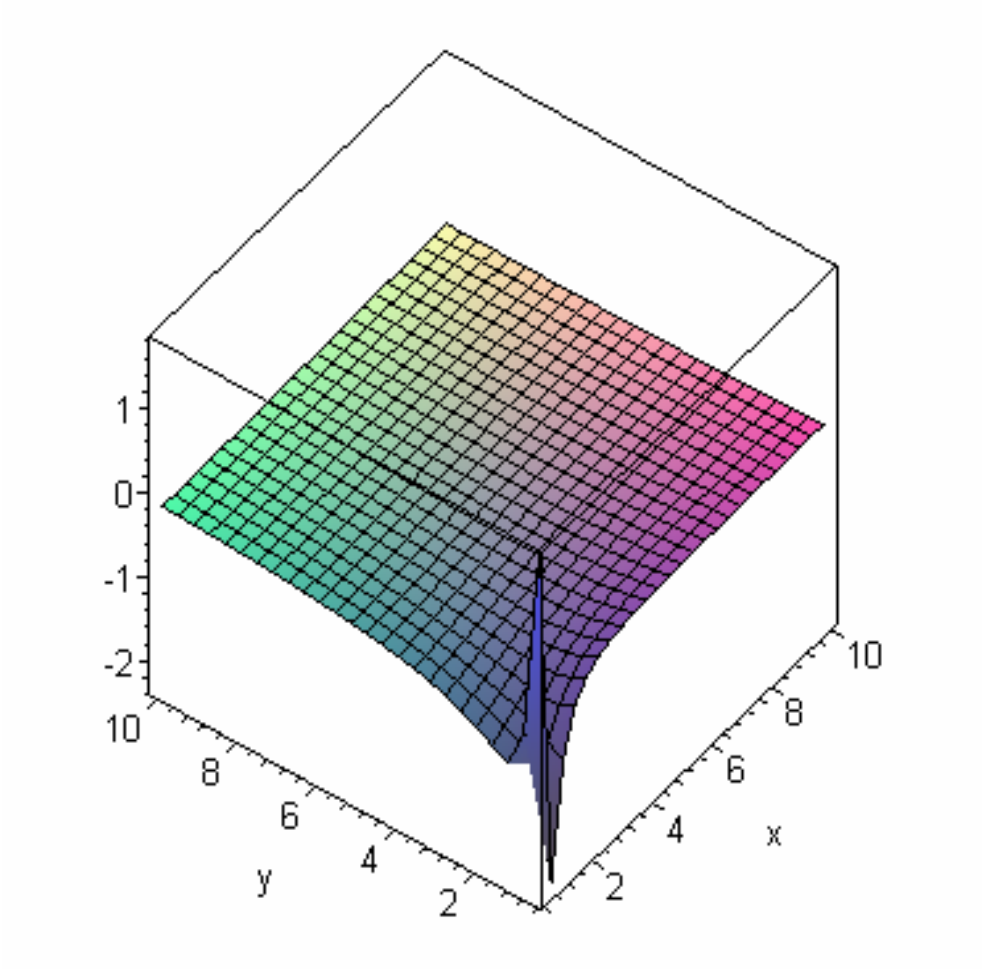

**Fig. 14. The 3D image of function,** $\mathrm{Re}\frac{\sqrt{\pi}}{z_0}\left(1+0.5\frac{1}{z_0^2}+0.75\frac{1}{z_0^4}\right)$, $(x=0.5,\ldots,10;\ y=0.5,\ldots,10)$ **(left).**

**Fig. 15. The 3D image of function** $\mathrm{Im}\frac{\sqrt{\pi}}{z_0}\left(1+0.5\frac{1}{z_0^2}+0.75\frac{1}{z_0^4}\right)$, $(x=0.5,\ldots,10;\ y=0.5,\ldots,10)$ (right).

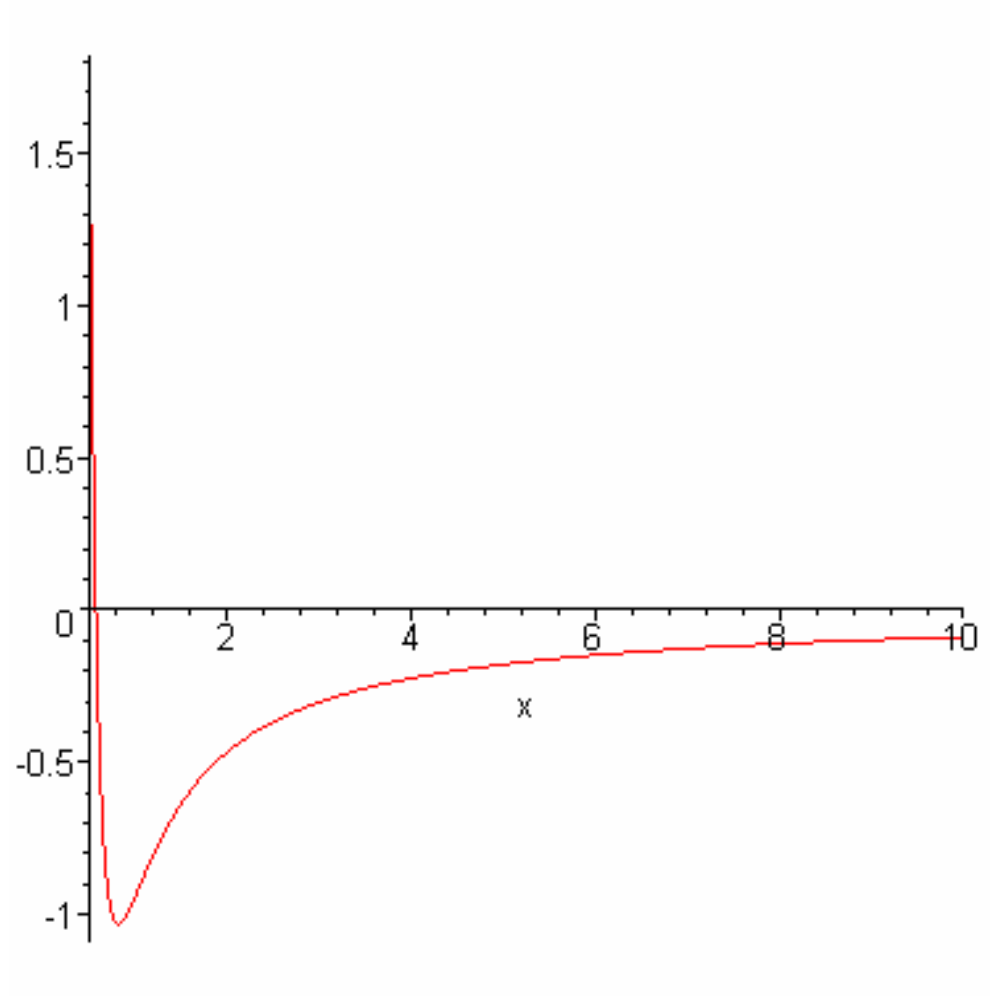

**Fig. 16. Sectional view of function** $\mathrm{Re}\frac{\sqrt{\pi}}{z_0}\left(1+0.5\frac{1}{z_0^2}+0.75\frac{1}{z_0^4}\right)$ $(x=0.5,\ldots,10;\, y=0.5,\ldots,10)$ **for line** $x=y$ **.**

What can be said about the accuracy of the first approximation $S_1=\frac{\sqrt{\pi}}{z_0}$ in comparison with the others $S_m$? The answer for this question can be received with the help of difference $H_m=S_m-S_1$. The following Figures 17 –20 correspond to difference $H_7=S_7-S_1$ and $H_{19}=S_{19}-S_1$.

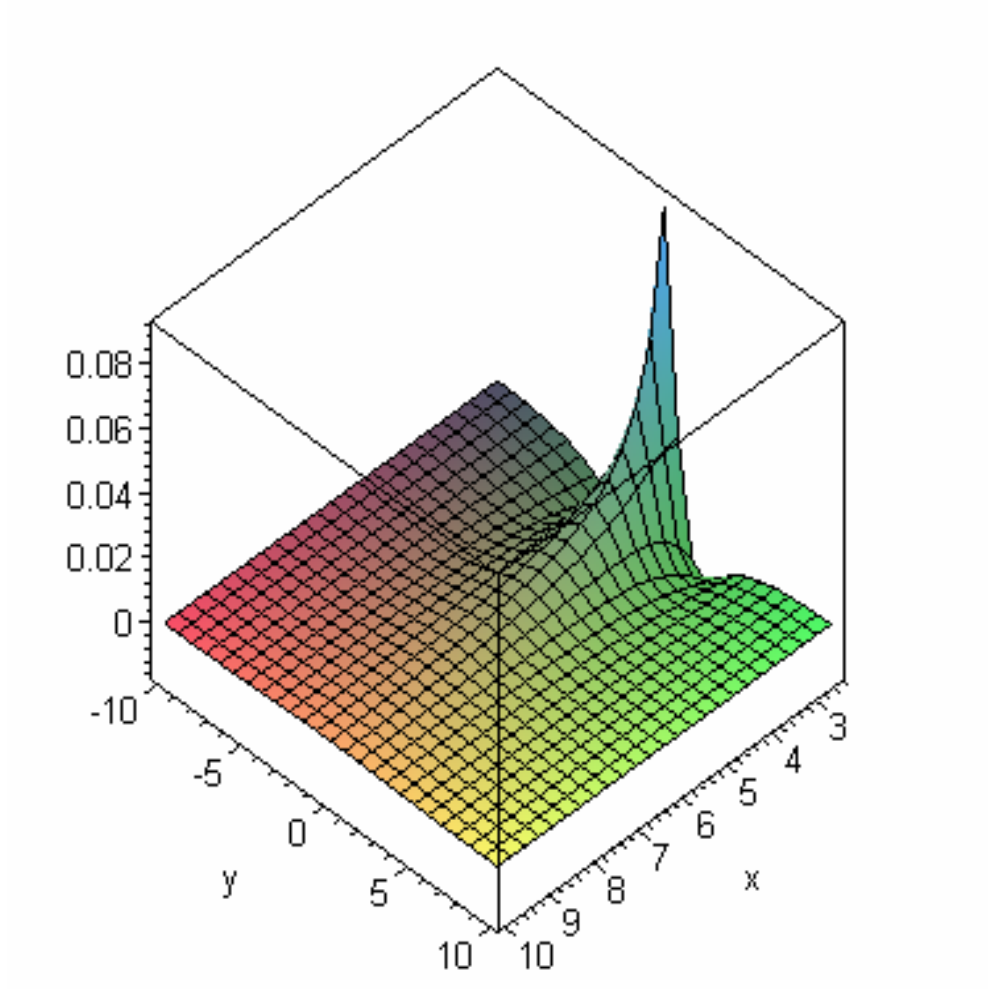

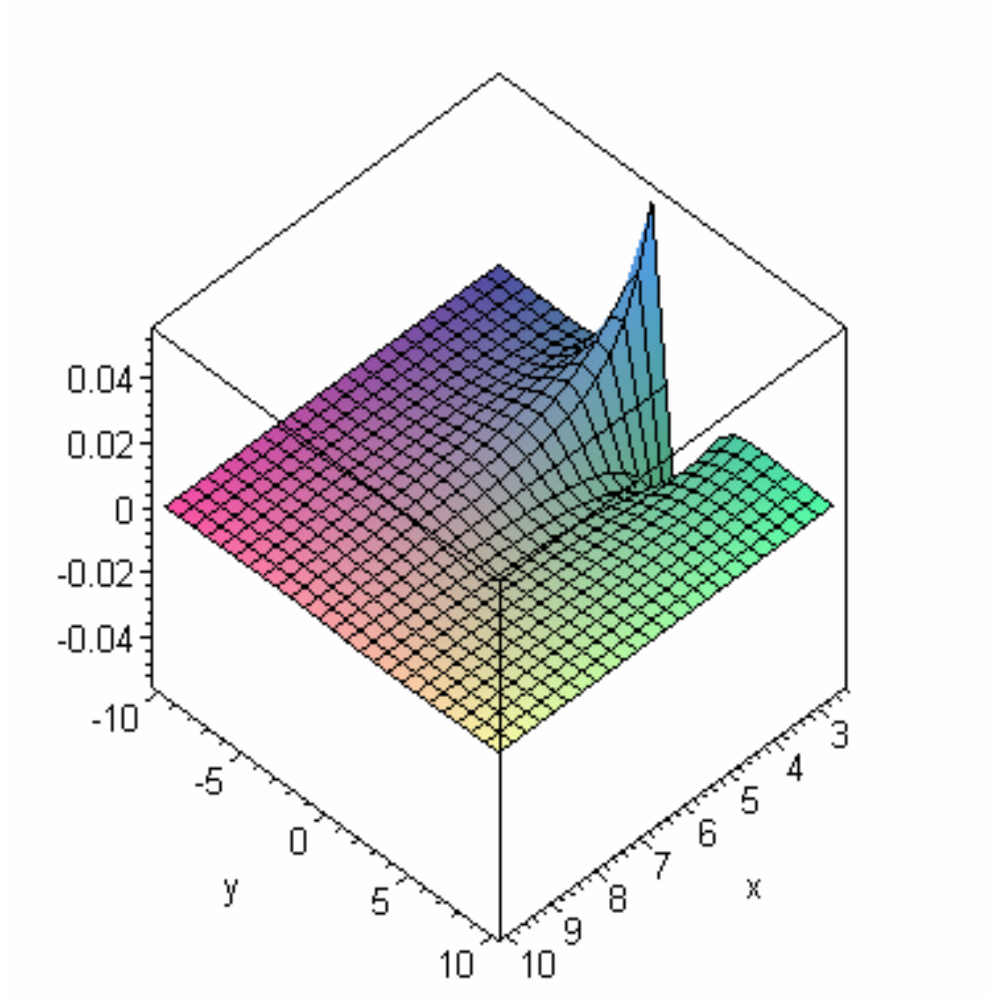

Fig. 17. The 3D image for the difference $\mathrm{Re}\,H_7=\mathrm{Re}(S_7-S_1)$, (left).
Fig. 18. The 3D image for the difference $\mathrm{Im}\,H_7=\mathrm{Im}(S_7-S_1)$, (right).

For the better observation of the differences $Re\,H_{19}=Re(S_{19}-S_1)$ and $Im\,H_{19}=Im(S_{19}-S_1)$ the region of heavy change of the functions are turned around the vertical axis.

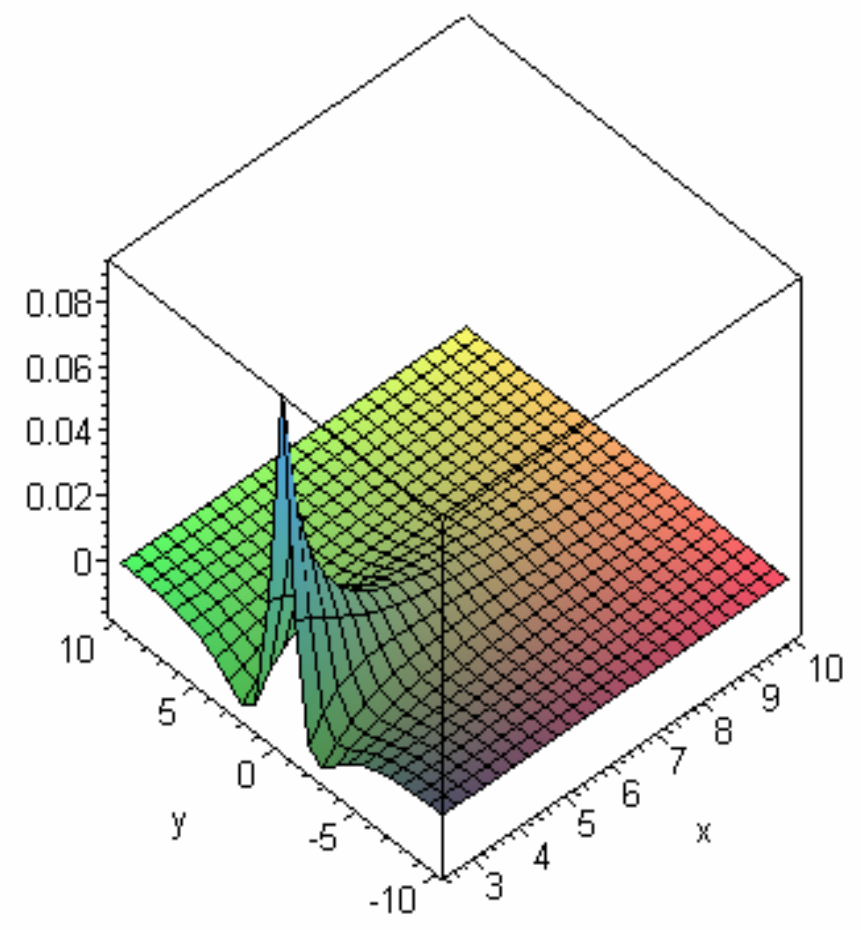

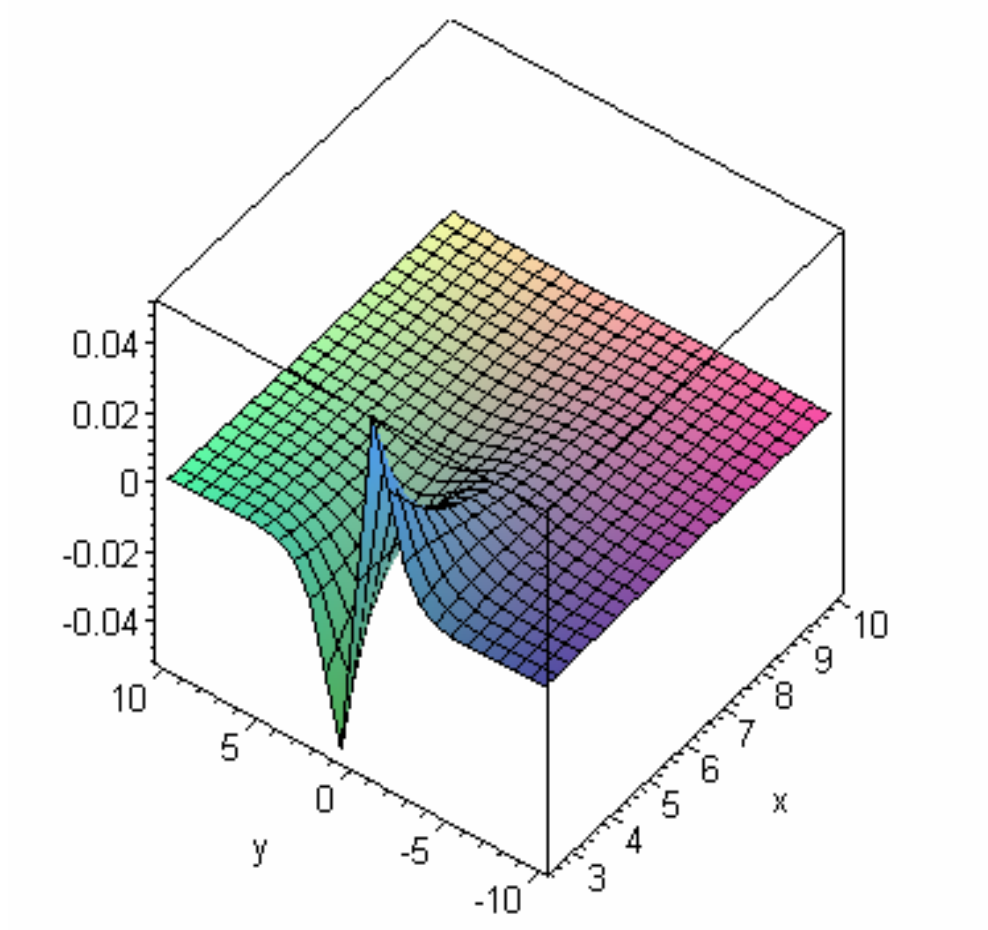

Fig. 19. The 3D image for the difference $\mathrm{Re}\,H_{19} = \mathrm{Re}(S_{19} - S_1)$, (left).
Fig. 20. The 3D image for the difference $\mathrm{Im}\,H_{19} = \mathrm{Im}(S_{19} - S_1)$, (right); $x = 2.5,...,10$; $y = -10,...,+10$.

The mathematical modeling attests about good convergence of successive approximations of the method of "large $z_0$" apart of the region near the image axes. From this point of view for accuracy investigation is sufficient to investigate the difference between the numerical evaluation of $L(z_0)$ and its presentation in the form of partial sum $S_1$. The corresponding difference is written as

$$H_{ex,1} = L(z_0) - S_1, \tag{2.27}$$

or

$$H_{ex,1} = \int_{-\infty}^{+\infty} \frac{e^{-t^2}}{z_0 - t} dt - \frac{\sqrt{\pi}}{z_0}. \tag{2.28}$$

(see Figures 21, 22)

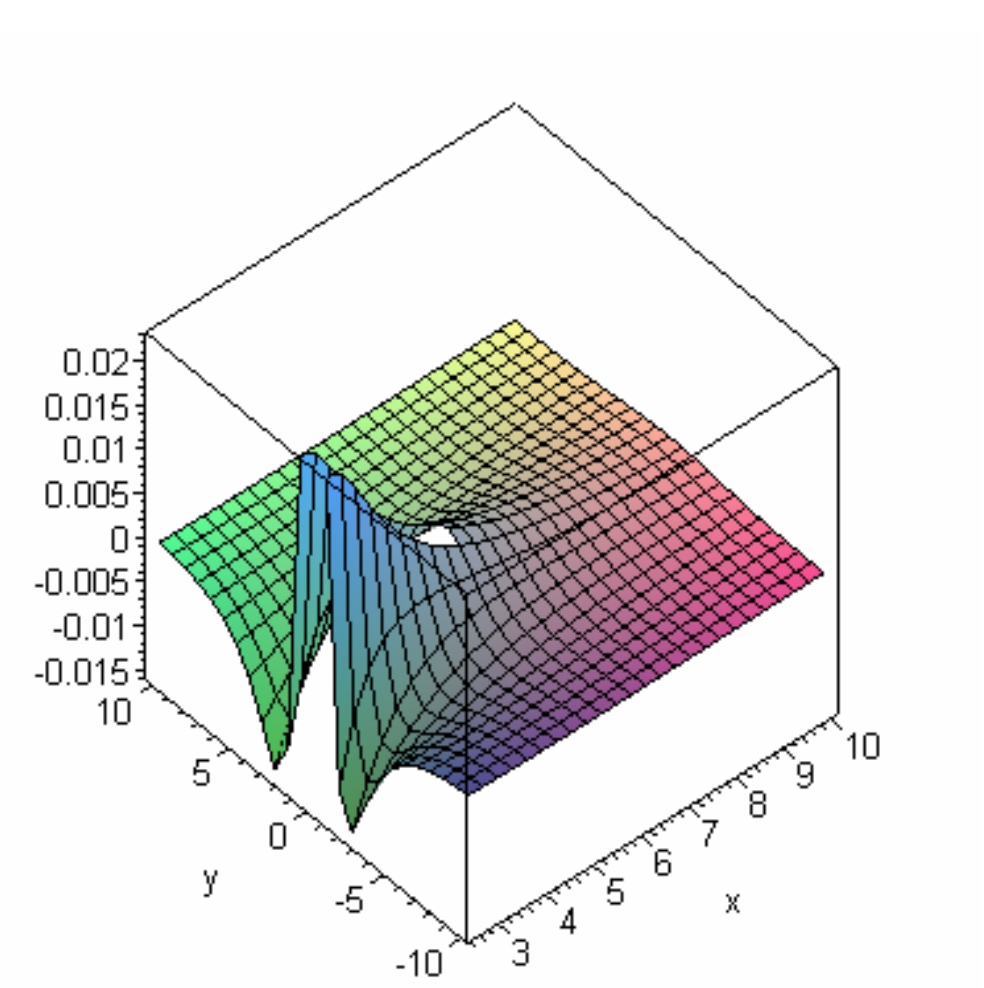

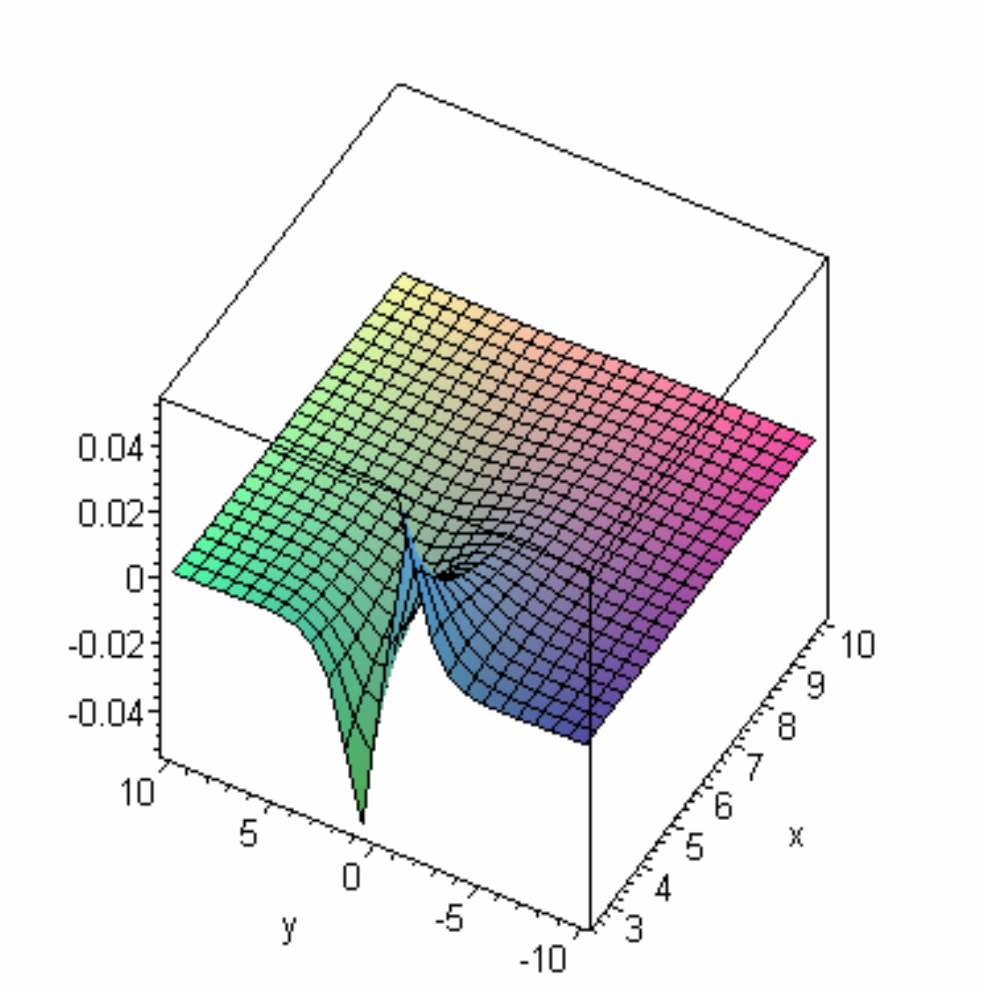

Fig. 21. The 3D image for the difference $\mathrm{Re}\,H_{ex,1}$, (left).
Fig. 22. The 3D image for the difference $\mathrm{Im}\,H_{ex,1}$, (right); $x = 2.5,...,10$; $y = -10,...,+10$.

**3. Estimation of the accuracy of Landau approximation.**

Let us remind the main steps of derivation of the Landau damping solution for dispersion equation (1.11) for the long wave limit case when $r_D^2 k^2 << 1$. Landau proposal for $L(z_0)$ approximation consists in combination of the sum of semi-residual (written for the upper plane) and three terms from the $L(z_0)$ series of the method of "large $z_0$":

$$L(z_0) = \int_{-\infty}^{+\infty} \frac{e^{-t^2}}{z_0 - t} dt = -i\pi e^{-z_0^2} + \frac{\sqrt{\pi}}{z_0}\left(1 + 0.5\frac{1}{z_0^2} + 0.75\frac{1}{z_0^4} + ...\right). \qquad (3.1)$$

Immediately should be noted that the first term in the right-hand-side of (3.1) coincides with

$$V = -\pi e^{-x^2} + 2\sqrt{\pi}y - 4\sqrt{\pi}xye^{-x^2}\int_0^x e^{s^2} ds \qquad (3.2)$$

from (2.20) for small y, but for upper side in spite of the solution is searching in the lower half plane.

After substitution (3.1) in (1.11) and transformation of (1.11) with the help of the nomenclature (1.10), (1.12) – (1.14) used above one obtains

$$i\omega\sqrt{\pi}\left(\frac{m_e}{2k_B T}\right)^{1.2} \frac{1}{r_D^2 k^3} e^{-\hat{\omega}^2} = -1 + \left(0.5\frac{1}{\hat{\omega}^2} + 0.75\frac{1}{\hat{\omega}^4} + ...\right)\frac{1}{r_D^2 k^2} \qquad (3.3)$$

or

$$i\omega\sqrt{\frac{\pi}{2}} \frac{\omega_e^2}{k^3\left(\frac{k_B T}{m_e}\right)^{3/2}} e^{-\hat{\omega}^2} = -1 + \frac{\omega_e^2}{\omega^2} + \frac{3}{4\hat{\omega}^4}\frac{1}{r_D^2 k^2}, \qquad (3.4)$$

where plasma frequency is introduced $\omega_e = \sqrt{k_B T / m_e} / r_D$.Using the relation

$$r_D^2 k^2 \hat{\omega}^4 = \frac{1}{k^2}\omega^4 \omega_e^{-2} \frac{m_e}{4k_B T} \qquad (3.5)$$

transform (3.4)

$$1 - \frac{\omega_e^2}{(\omega' + i\omega'')^2} - \frac{3\omega_e^2}{(\omega' + i\omega'')^4} k^2\left(\frac{k_B T}{m_e}\right) + i(\omega' + i\omega'')\sqrt{\frac{\pi}{2}} \frac{\omega_e^2}{k^3\left(\frac{k_B T}{m_e}\right)^{3/2}} e^{-(\hat{\omega}' + i\hat{\omega}'')^2} = 0. \qquad (3.6)$$

If

$$|\omega''| << \omega', \qquad (3.7)$$

(it is valid for Landau solution), then separation of real and imaginary parts leads to relations

$$1 - \frac{\omega_e^2}{\omega'^2} = 0, \qquad (3.8)$$

(real part of (3.6) in the first approximation) or

$$\omega' = \omega_e. \qquad (3.9)$$

Formally from Landau solution follows also for real part

$$1 - \frac{\omega_e^2}{\omega'^2} - \frac{3\omega_e^2}{\omega'^4} k^2\left(\frac{k_B T}{m_e}\right) = 0 \qquad (3.10)$$

or

$$\omega'^2 = \omega_e^2 + 3k^2\left(\frac{k_B T}{m_e}\right). \qquad (3.11)$$

Imaginary part of (3.6) has the form

$$\omega'\sqrt{\frac{\pi}{2}}\frac{\omega_e^2}{k^3\left(\frac{k_B T}{m_e}\right)^{3/2}}e^{-\hat{\omega}'^2}=-\frac{\omega_e^2}{\omega'^4}2\omega'\omega'' \tag{3.12}$$

For damping oscillations the decrement $\gamma$ can be introduced as

$$\gamma=-\omega'' \tag{3.13}$$

and

$$\gamma=\sqrt{\frac{\pi}{8}}\frac{\omega_e^4}{k^3\left(\frac{k_B T}{m_e}\right)^{3/2}}e^{-\hat{\omega}_e'^2}, \tag{3.14}$$

where

$$\hat{\omega}_e'^2=\frac{1}{2k^2 r_D^2} \tag{3.15}$$

It leads to the standard form for the Landau formula

$$\gamma=\sqrt{\frac{\pi}{8}}\frac{\omega_e^4}{k^3\left(\frac{k_B T}{m_e}\right)^{3/2}}\exp\left[-\frac{1}{2k^2 r_D^2}\right]. \tag{3.16}$$

In the long wave case Landau formula leads to very small $\gamma$, this fact (see (3.7)) was used beforehand by the transformations in Landau approximation.

Realize now the direct numerical integration for $L(z_0)$ near the real axis for analysis of the Landau solution. Fig. 23 contains the integral surface of $\mathrm{Re}\,L(z_0)$ near real axis $(x=-10,...,+10;\ y=-0.01,...,-0.001)$, and Fig. 24 - – the integral surface $\mathrm{Im}\,L(z_0)$ also near real axis $(x=-10,...,+10;\ y=-0.01,...,-0.001)$. What is the difference between these results and the Landau approximation for $L(z_0)$ (see also (3.1))? The answers for these questions can be found from the Figures 25, 26, which reproduce the difference $\mathrm{Im}\left[L(z_0)+\pi i e^{-z_0^2}-\frac{\sqrt{\pi}}{z_0}\right]$ near the real axis for the domains $(x=0,...,+5;\ y=-0.01,...,-0.001)$ and $(x=2,...,+5;\ y=-0.01,...,-0.001)$ correspondingly:

1. The results for small $x$ are absolutely inadequate.
2. For $x=2$ the imaginary part of difference between the exact solution and the Landau approximation is about of 0.12. It seems not too bad – the approximation of “large $z_0$” begins to work. But strict application of combination of semi-residual and the approximation of “large $z_0$” for lower half-plane leads to changing sign in front of the exponential term in the right-hand-side of Eq. (3.1) (see strict results (2.21), (2.22)).
3. The correction of sign corresponds to Fig. 27. The correction of sign (which leads to the integral surface $\mathrm{Im}\left[L(z_0)-\pi i e^{-z_0^2}-\frac{\sqrt{\pi}}{z_0}\right]$) gives the much better approximation near the real axis $(x=2,...,+5;\ y=-0.01,...,-0.001)$. The corresponding difference for $x=2$ does not exceed 0.0005.
4. As it was demonstrated even the first term of the approximation of “large $z_0$” gives good approximation of $L(z_0)$ for large $x$. But, because Landau used three terms of the approximation of "large $z_0$", the following Fig. 28, Fig. 29 demonstrate the difference $\operatorname{Im}\left[L(z_0) - \pi i e^{-z_0^2} - \frac{\sqrt{\pi}}{z_0}\left(1 + \frac{0.5}{z_0^2} + \frac{0.75}{z_0^4}\right)\right]$ in vicinity of real axis $(x = 2,\dots,+5;\ \ y = -0.01,\dots,-0.001)$ by the right choice of sign, and for integral surface $\operatorname{Im}\left[L(z_0) + \pi i e^{-z_0^2} - \frac{\sqrt{\pi}}{z_0}\left(1 + \frac{0.5}{z_0^2} + \frac{0.75}{z_0^4}\right)\right]$ for the Landau approximation.

5. *The right correction of the sign leads to liquidation the Landau damping effect, moreover to* $\hat{\omega}'' > 0$ *and to reinforcement of oscillations.*

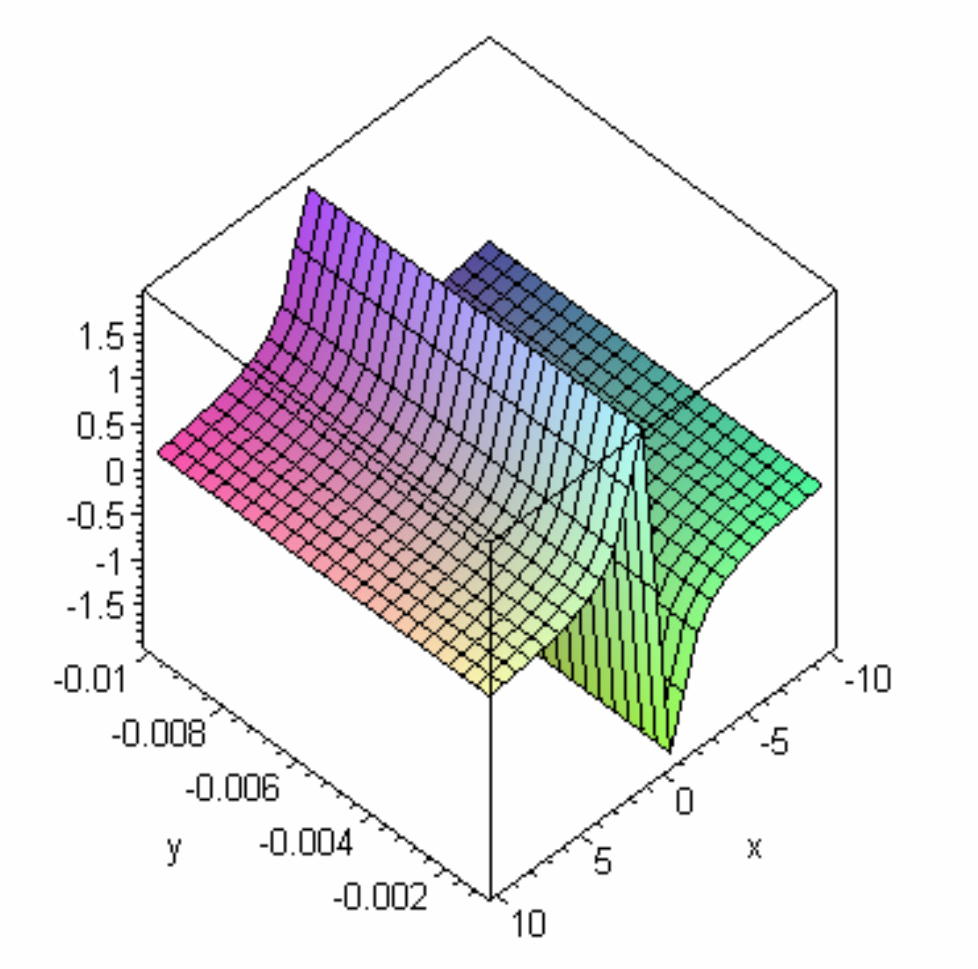

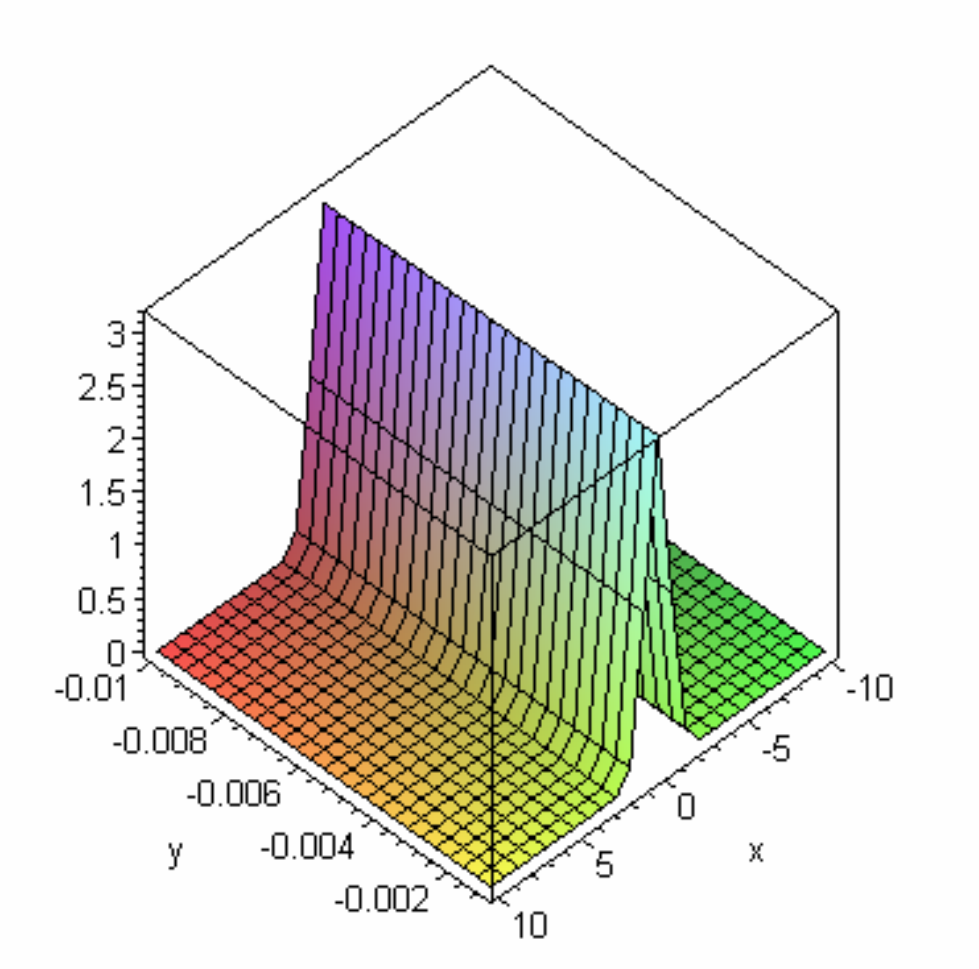

Fig. 23. The integral surface of $\operatorname{Re} L(z_0)$ near real axis $(x = -10,\dots,+10;\ \ y = -0.01,\dots,-0.001)$, (left).
Fig. 24. The integral surface of $\operatorname{Im} L(z_0)$ near real axis $(x = -10,\dots,+10;\ \ y = -0.01,\dots,-0.001)$, (right).

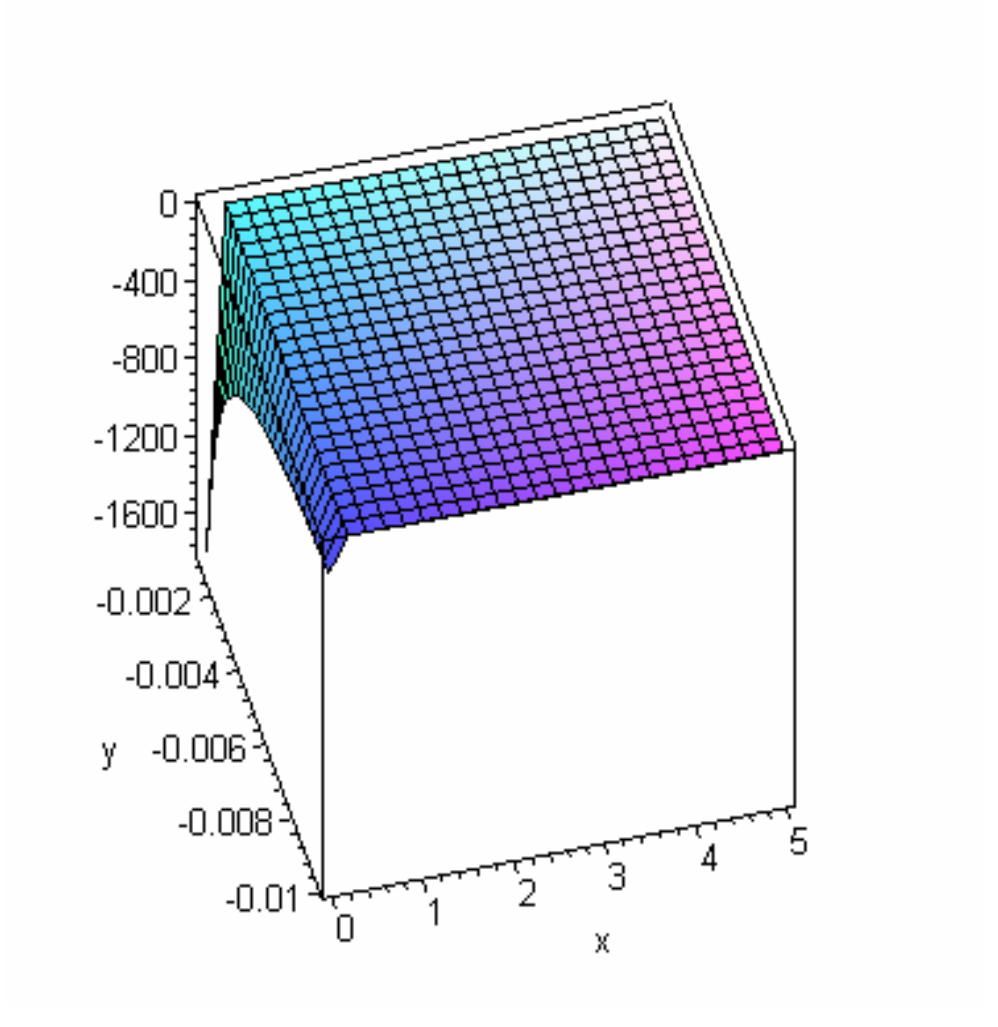

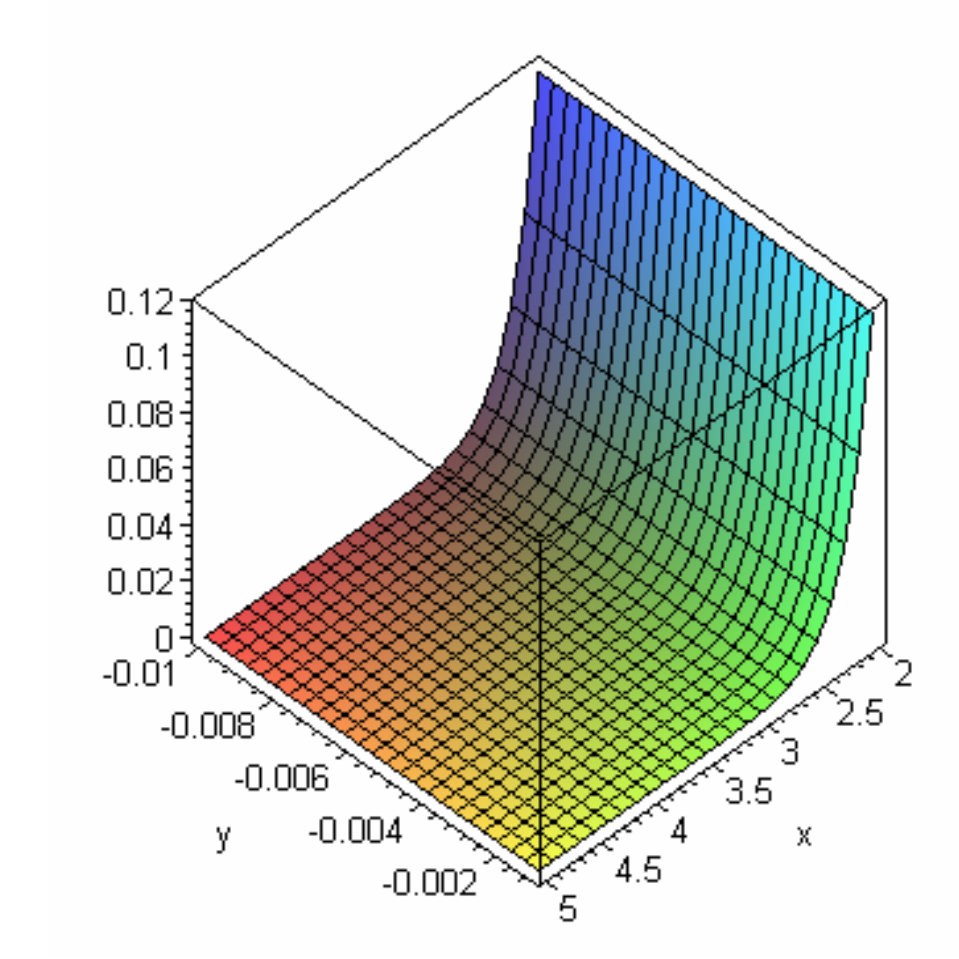

Fig. 25. The integral surface of $\operatorname{Im}\left[L(z_0) + \pi i e^{-z_0^2} - \frac{\sqrt{\pi}}{z_0}\right]$ near real axis $(x = 0,\dots,+5;\ \ y = -0.01,\dots,-0.001)$, (left).

Fig. 26. The integral surface of $\mathrm{Im}\left[L(z_0)+\pi i e^{-z_0^2}-\frac{\sqrt{\pi}}{z_0}\right]$ near real axis $(x=2,\dots,+5;\ \ y=-0.01,\dots,-0.001)$, (right).

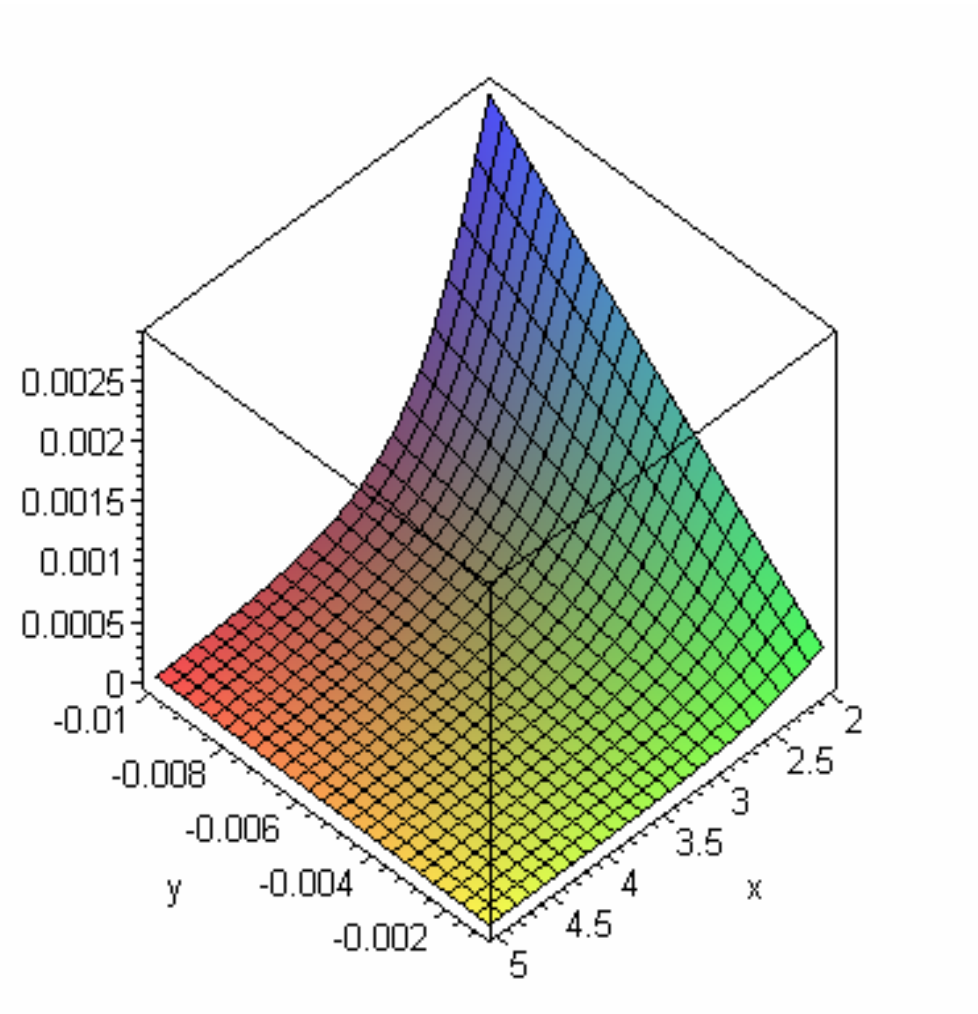

Fig. 27. The integral surface of $\mathrm{Im}\left[L(z_0)-\pi i e^{-z_0^2}-\frac{\sqrt{\pi}}{z_0}\right]$ near real axis $(x=2,\dots,+5;\ \ y=-0.01,\dots,-0.001)$.

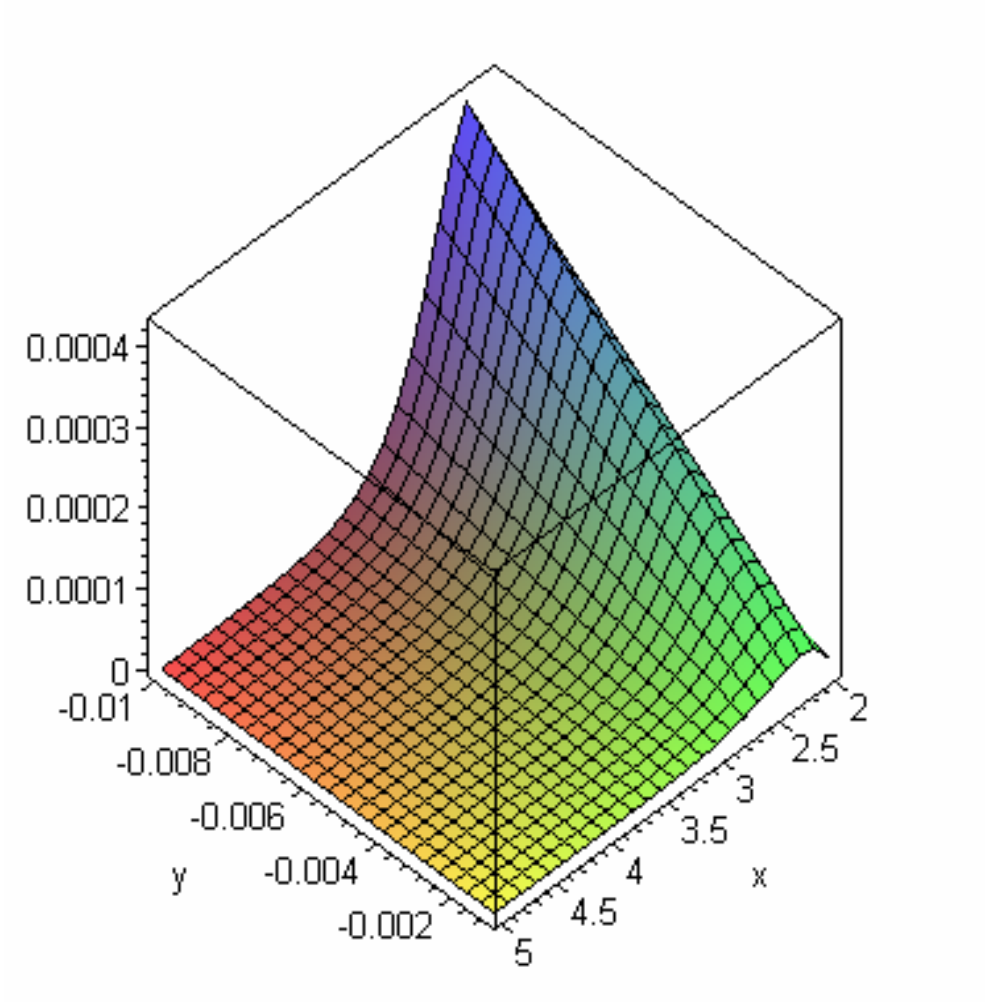

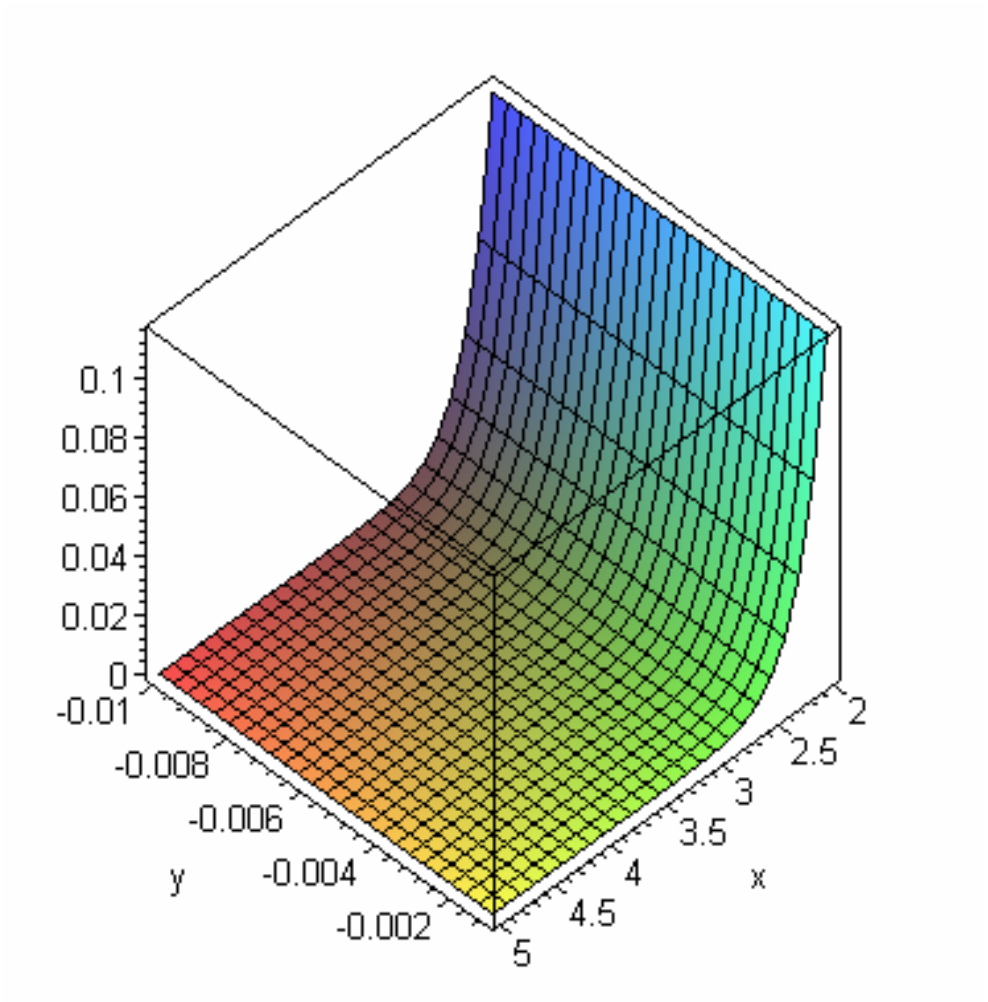

Fig. 28. The integral surface of $\mathrm{Im}\left[L(z_0)-\pi i e^{-z_0^2}-\frac{\sqrt{\pi}}{z_0}\left(1+\frac{0.5}{z_0^2}+\frac{0.75}{z_0^4}\right)\right]$ near real axis $(x=2,\dots,+5;\ \ y=-0.01,\dots,-0.001)$ (left).

Fig. 29. The integral surface of $\mathrm{Im}\left[L(z_0)+\pi i e^{-z_0^2}-\frac{\sqrt{\pi}}{z_0}\left(1+\frac{0.5}{z_0^2}+\frac{0.75}{z_0^4}\right)\right]$ near real axis $(x=2,\dots,+5;\ \ y=-0.01,\dots,-0.001)$. The difference for the Landau approximation (right).

As we see the Landau approximation for the complex integral function $L(z)$ leads to the mathematical contradictions in spite of clear physical sense of effect and the experimental confirmations.

*So called "Landau rule" should be considered as the additional condition implied for physical system.*

**4. Alternative analytical solutions of the Landau – Vlasov dispersion equation**

From the theory of complex variables is known Cauchy's integral formula: if the function $f(z)$ is analytic inside and on a simple closed curve *C*, and $z_0$ is any point inside C, then

$$f(z_0) = -\frac{1}{2\pi i}\oint_C \frac{f(z)}{z_0 - z}dz \tag{4.1}$$

where *C* is traversed in the positive (counterclockwise) sense.

Let *C* be the boundary of a simple closed curve placed in lower half plane (for example a semicircle of radius *R*) with the corresponding element of real axis, $z_0$ is an interior point. As usual after adding to this semicircle a cross-cut connecting semicircle *C* with the interior circle (surrounding $z_0$) of the infinite small radius for analytic $f(z)$ the formula takes place

$$\oint_C \frac{f(z)}{z_0 - z}dz = -\int_{-R}^{R} \frac{f(\tilde{x})}{z_0 - \tilde{x}}d\tilde{x} + \int_{C_R} \frac{f(z)}{z_0 - z}dz + 2\pi i f(z_0), \tag{4.2}$$

because the integrals along cross-cut cancel each other, ( $z = \tilde{x} + i\tilde{y}$ ).

Analogically for upper half plane

$$\oint_C \frac{f(z)}{z_0 - z}dz = \int_{-R}^{R} \frac{f(\tilde{x})}{z_0 - \tilde{x}}d\tilde{x} + \int_{C_R} \frac{f(z)}{z_0 - z}dz + 2\pi i f(z_0). \tag{4.3}$$

The formulae (4.2), (4.3) could be used for calculation (including the case $R \to \infty$) of the integrals along the real axis with the help of the residual theory *for arbitrary* $z_0$ if analytical function $f(z)$ satisfies the special conditions of decreasing by $R \to \infty$.

Let us consider now integral $\int_{C_R} \frac{e^{-z^2}}{z_0 - z}dz$. Generally speaking for function $f(z) = e^{-z^2}$ Cauchy's conditions are not satisfied. Really for a point $z = \tilde{x} + i\tilde{y}$ this function $f(z) = e^{\tilde{y}^2 - \tilde{x}^2}\left[\cos(2\tilde{x}\tilde{y}) - i\sin(2\tilde{x}\tilde{y})\right]$ and by $|\tilde{y}| > |\tilde{x}|$ the function is growing for this part of $C_R$.

*But from physical point of view in the linear problem of interaction of individual electrons only with waves of potential electric field the natural assumption can be introduced that solution depends only of concrete* $z_0 = \hat{\omega}' + i\hat{\omega}''$, *but does not depend of another possible modes of oscillations in physical system.*

It can be realized only if the calculations do not depend of choosing of contour $C_R$. This fact leads to the additional conditions, for lower half plane

$$\int_{-\infty}^{\infty} \frac{f(\tilde{x})}{z_0 - \tilde{x}}d\tilde{x} = 2\pi i f(z_0), \tag{4.4}$$

and for upper half plane

$$\int_{-\infty}^{\infty} \frac{f(\tilde{x})}{z_0 - \tilde{x}}d\tilde{x} = -2\pi i f(z_0). \tag{4.5}$$

It can be shown (using Eqs. (4.2) and (3.1)) that Landau approximation contains in implicit form restrictions (near the vicinity of $\tilde{x}$-axis) in the contour $C$ choosing of the same physical sense mentioned above.

The question arises, is it possible to find solutions of the equation (1.11) by the restriction (4.4)? In the following will be shown that the condition (4.4), (4.5) together with (1.11) lead to the discrete spectrum of $z_0 = \hat{\omega}' + i\hat{\omega}''$ and from physical point of view condition (4.4) can be considered as condition of quantization.

Substitution for example (4.4) into (1.11) gives the result

$$2\sqrt{\pi} e^{-z_0^2} = -i \frac{\beta}{z_0}, \tag{4.6}$$

where $\beta = 1 + k^2 r_D^2$. The condition (4.5) leads to the formal changing of the sign in front of $\beta$. The following construction of solution will depend on $\beta^2$, has the universal character but needs to corresponding choosing of solutions for the low and upper planes. Write down (4.6) via complex frequencies

$$2\sqrt{\pi} e^{-\hat{\omega}'^2 + \hat{\omega}''^2 - 2\hat{\omega}'\hat{\omega}'' i} = -i \frac{\beta}{\hat{\omega}' + i\hat{\omega}''} \tag{4.7}$$

and separate the real and imaginary parts of equation. Real part

$$-\frac{\beta}{2\sqrt{\pi}} e^{\hat{\omega}'^2 - \hat{\omega}''^2} \cos(2\hat{\omega}'\hat{\omega}'') = \hat{\omega}'', \tag{4.8}$$

imaginary part

$$-\frac{\beta}{2\sqrt{\pi}} e^{\hat{\omega}'^2 - \hat{\omega}''^2} \sin(2\hat{\omega}'\hat{\omega}'') = -\hat{\omega}'. \tag{4.9}$$

After dividing of equation (4.8) on (4.9), one obtains

$$\hat{\omega}' \cos(2\hat{\omega}'\hat{\omega}'') + \hat{\omega}'' \sin(2\hat{\omega}'\hat{\omega}'') = 0. \tag{4.10}$$

After introducing notation

$$\alpha = 2\hat{\omega}'\hat{\omega}'', \tag{4.11}$$

the following system of transcendent equations takes place

$$-\frac{\beta}{4\sqrt{\pi}} e^{\hat{\omega}'^2 - \hat{\omega}''^2} \sin 2\alpha = \hat{\omega}'' \sin\alpha, \tag{4.12}$$

$$\hat{\omega}' \cos\alpha + \hat{\omega}'' \sin\alpha = 0. \tag{4.13}$$

Using the relations

$$\hat{\omega}'^2 = -\frac{1}{2}\alpha tg\alpha, \quad \hat{\omega}''^2 = -\frac{1}{2}\alpha ctg\alpha, \quad \hat{\omega}'^2 - \hat{\omega}''^2 = \alpha ctg 2\alpha. \tag{4.14}$$

we find

$$e^{\alpha ctg 2\alpha} \sin 2\alpha = -\frac{4\sqrt{\pi}}{\beta} \hat{\omega}'' \sin\alpha, \tag{4.15}$$

and after squaring of both parts of (4.15)

$$e^{2\alpha ctg 2\alpha} \sin^2 2\alpha = \frac{16\pi}{\beta^2} \hat{\omega}''^2 \sin^2\alpha \tag{4.16}$$

As it was mentioned, this equation does not depend on the sign in front of parameter $\beta$. Using once more (4.14), we find

$$e^{2\alpha ctg 2\alpha}\sin 2\alpha = -\frac{4\pi}{\beta^2}\alpha . \quad (4.17)$$

After introducing notation

$$\sigma = -2\alpha \quad (4.18)$$

the dispersion equation takes the finalized form

$$e^{\sigma ctg\sigma}\sin\sigma = -\frac{2\pi}{\beta^2}\sigma . \quad (4.19)$$

The exact solution of equation (4.19) can be found with the help of the $W$-function of Lambert

$$\sigma_n = \mathrm{Im}\left[ W_n\left(\frac{\beta^2}{2\pi}\right)\right], \quad (4.20)$$

frequencies $\hat{\omega}'_n, \hat{\omega}''_n$ are (see (1.14), (4.14), (4.18))

$$\omega'_n = k\sqrt{-\frac{k_B T}{2m_e}\sigma_n \, tan\frac{\sigma_n}{2}} \, , \; \omega''_n = -k\sqrt{-\frac{k_B T}{2m_e}\sigma_n \, cot\frac{\sigma_n}{2}} \, . \quad (4.21)$$

In asymptotic for large entire positive $n$

$$\sigma_n = \left(n+\frac{1}{2}\right)\pi \, , \; \hat{\omega}'_n = \frac{\sqrt{\sigma_n}}{2} = \frac{1}{2}\sqrt{\pi\left(n+\frac{1}{2}\right)} \, , \; \hat{\omega}''_n = -\frac{\sqrt{\sigma_n}}{2} = -\frac{1}{2}\sqrt{\pi\left(n+\frac{1}{2}\right)} . \quad (4.22)$$

The exact solution for the $n$ – th discrete solution from the spectrum of oscillations:

$$\hat{\omega}_n = \frac{1}{2}\sqrt{-\mathrm{Im}\left[W_n\left(\frac{\left(1+r_D^2 k^2\right)^2}{2\pi}\right)\right]\mathrm{tg}\left[\frac{1}{2}\mathrm{Im}\left[W_n\left(\frac{\left(1+r_D^2 k^2\right)^2}{2\pi}\right)\right]\right]} - \frac{i}{2}\sqrt{-\mathrm{Im}\left[W_n\left(\frac{\left(1+r_D^2 k^2\right)^2}{2\pi}\right)\right]\mathrm{ctg}\left[\frac{1}{2}\mathrm{Im}\left[W_n\left(\frac{\left(1+r_D^2 k^2\right)^2}{2\pi}\right)\right]\right]} . \quad (4.23)$$

The square of the oscillation frequency of plasma waves, $\omega_n^{'2}$ is proportional to the wave energy. Hence, the energy of plasma waves is quantized, and as $n$ grows we have the asymptotic expression

$$\hat{\omega}_n^{'2} = \frac{\pi}{4}\left(n+\frac{1}{2}\right) \quad (4.24)$$

and the squares of possible dimensionless frequencies become equally spaced:

$$\hat{\omega}_{n+1}^{'2} - \hat{\omega}_n^{'2} = \frac{\pi}{4} . \quad (4.25)$$

Figures 30 and 31 reflect the result of calculations for 200 discrete levels for low complex half plane. For high levels this spectrum contains many very close practically straight lines, which human eyes can perceive as background. Moreover plotter from the technical point of view has no possibility to reflect the small curvature of lines approximating this curvature as a long step. My suggestion is to turn this shortcoming into merit in explication of topology of high quantum levels in quantum systems.

Really, extremely interesting that this (from the first glance) grave shortcoming of plotters lead to the automatic construction of approximation for derivatives $d(r_D k)/d\hat{\omega}'$ and $d(r_D k)/d\hat{\omega}''$. This effect has no attitude to the mathematical programming. You can see this very complicated topology of curves including the spectrum of the bell-like dispersion curves in Fig. 30 and Fig. 31, which also form the discrete spectrum.

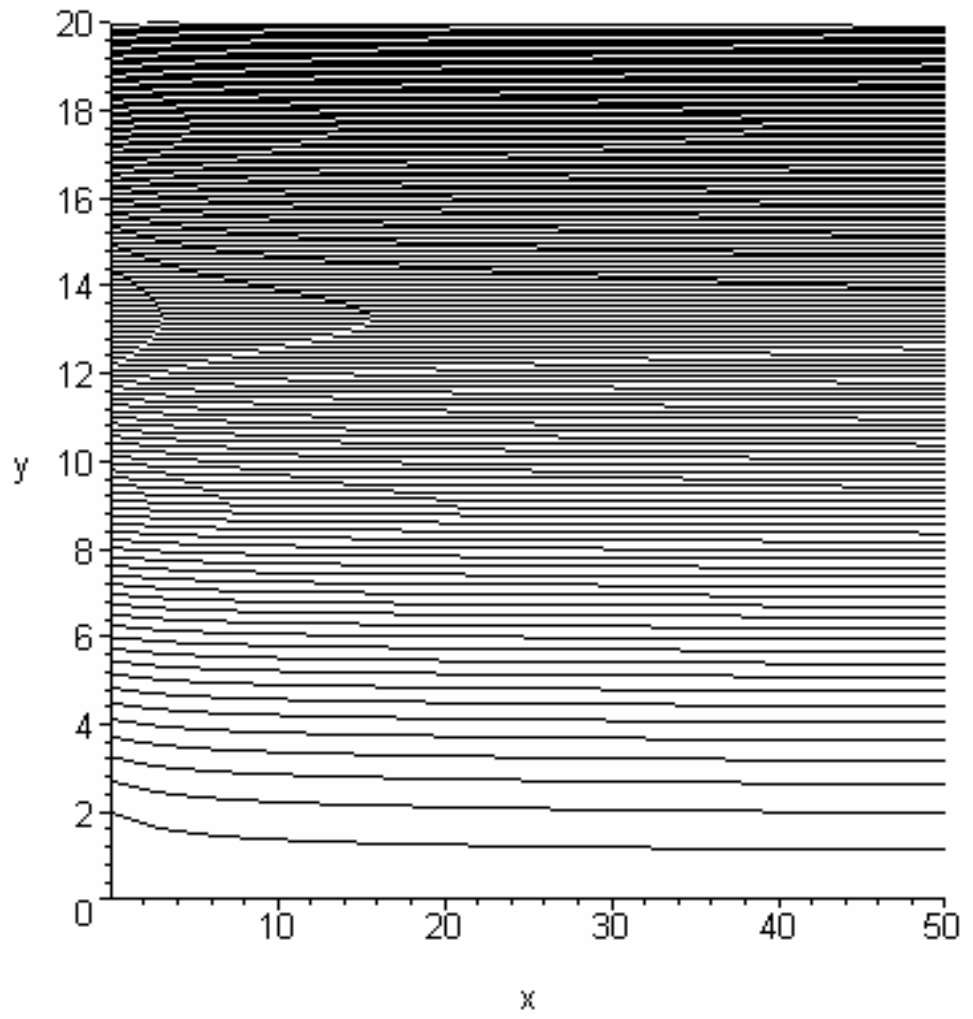

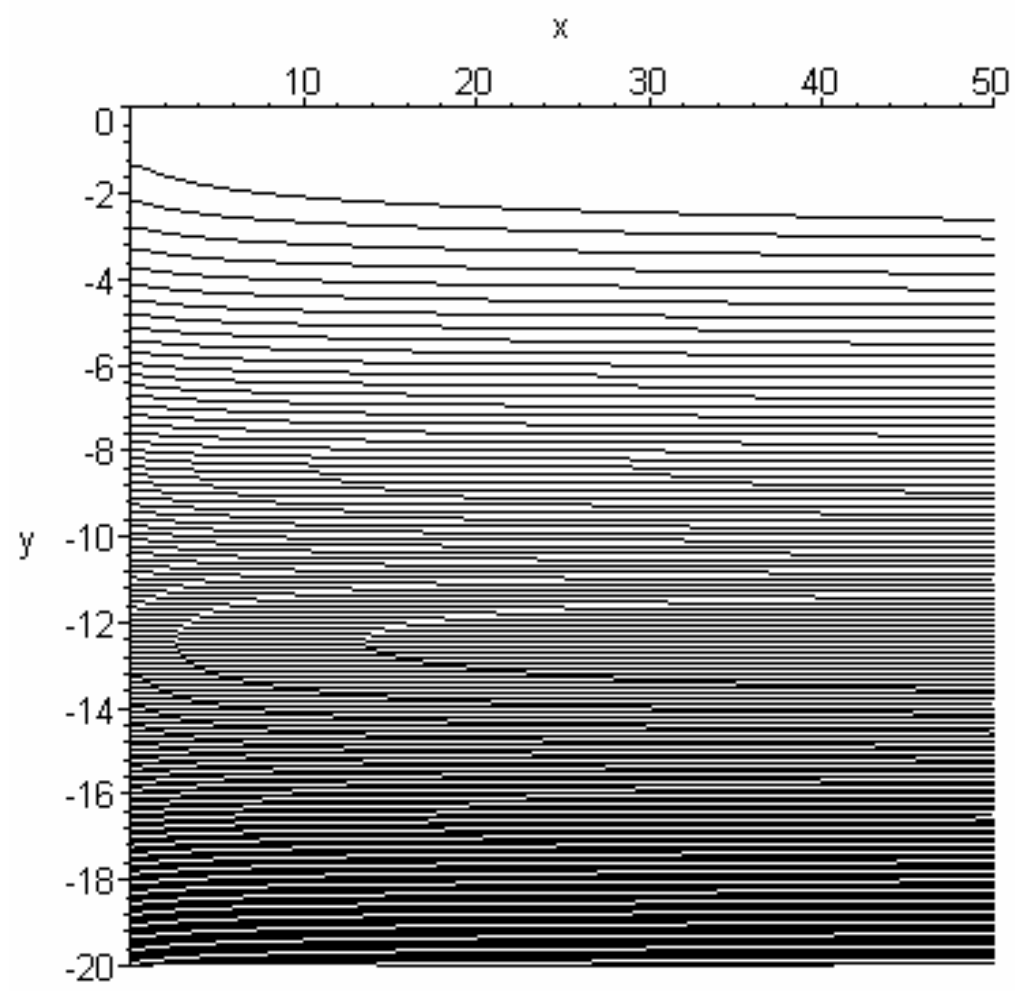

Fig. 30. The dimensionless frequency $\hat{\omega}'$ ($y$ axes) versus parameter $r_D k$ ($x$ axes); (left).
Fig. 31. The dimensionless frequency $\hat{\omega}''$ ($y$ axes) versus parameter $r_D k$ ($x$ axes); (right).

But mentioned derivatives can be written in the form

$$\frac{d(r_D k)}{d\hat{\omega}'} = \frac{dk}{d(\omega'/k)} \frac{k_B T\sqrt{2}}{m_e \omega_e} = \frac{dk}{dv_\phi} \frac{k_B T\sqrt{2}}{m_e \omega_e}, \tag{4.26}$$

where $v_\phi$ is the wave phase velocity. Then

$$\frac{dv_\phi}{d\lambda} = -\frac{2\pi}{\lambda^2} \frac{k_B T\sqrt{2}}{m_e \omega_e} \left[\frac{d(r_D k)}{d\hat{\omega}'}\right]^{-1} \tag{4.27}$$

or

$$\frac{d(r_D k)}{d\hat{\omega}'} = \frac{dk}{dv_\phi} \frac{k_B T}{e\sqrt{2\pi\rho_e}}, \tag{4.28}$$

$$\frac{dv_\phi}{d\lambda} = -\frac{2\pi}{\lambda^2} \frac{k_B T}{e\sqrt{2\pi\rho_e}} \left[\frac{d(r_D k)}{d\hat{\omega}'}\right]^{-1}, \tag{4.29}$$

where $\rho_e = m_e n_e$. Very complicated topology of the dispersion curves can be revealed by the construction of dependence $\hat{\omega}''/\hat{\omega}_e$ versus $r_D k$, (Fig. 32). By the curves compression we can see very complicated topology until the intermediate area passes over the black domain where (in chosen scale) the curves cannot be observed separately. Enlarging of scaling shows that the complicated curves topology exists also in the black domain. Then Fig. 30 – 32 can be used for understanding of the future development of events in physical system after the initial linear stage. For example Fig. 30 shows the discrete set of frequencies which vicinity corresponds to passing over from abnormal to normal dispersion (for example, by $\hat{\omega}' \sim 9$) for discrete systems of $r_D k$. Of course the non-linear stage needs the special investigation with using of another methods including the method of direct mathematical modeling [9-12].

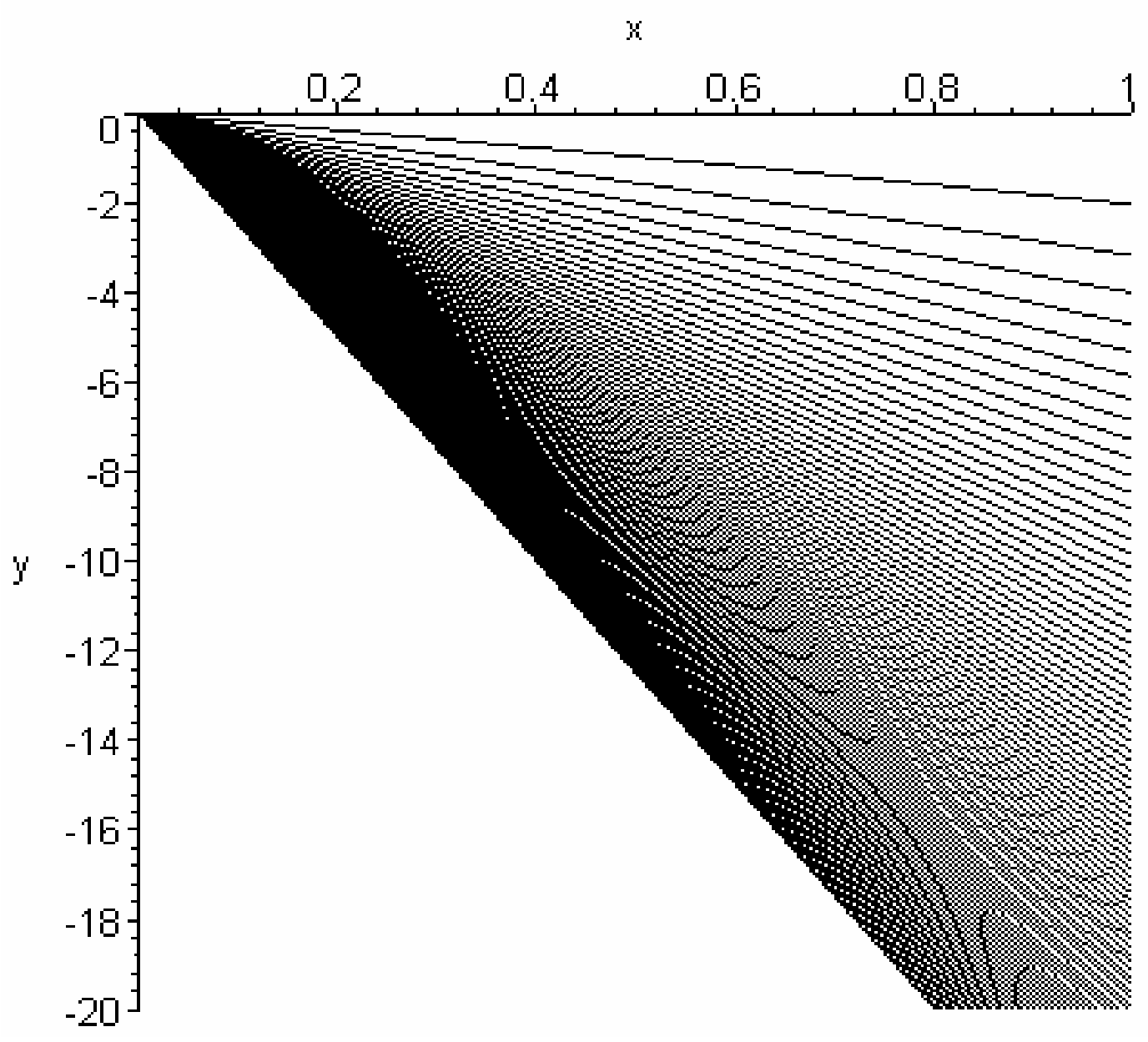

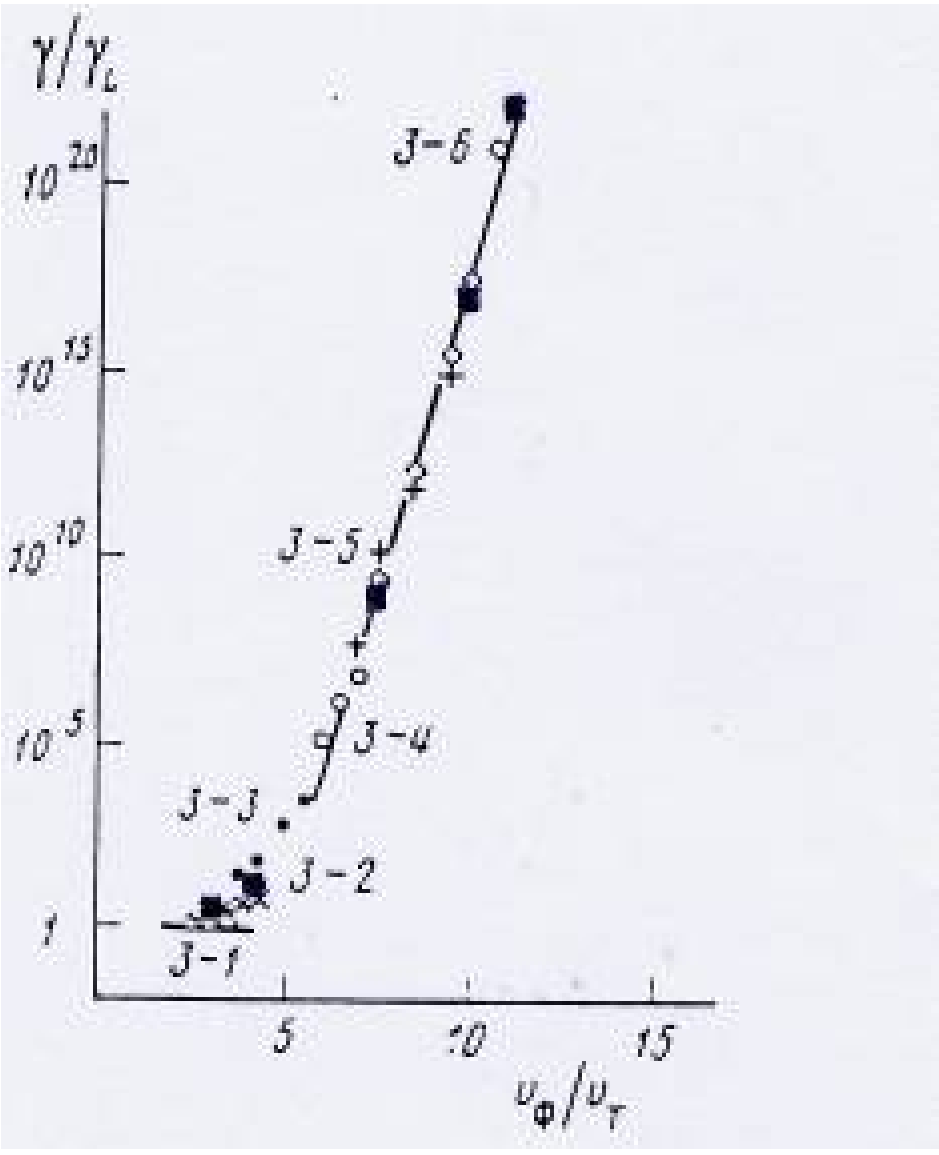

Fig. 32. Dependence $\hat{\omega}'' / \hat{\omega}_e$ ($y$ axes) on $r_D k$ ($x$ axes), (left).

Fig. 33. Dependence $\gamma / \gamma_L$ ($y$ axes) on $v_\phi / v_T$, ($x$ axes), (right).

We now proceed to compare the theoretical results with those of the computer experiment. Extensive simulations of the attenuation of Langmuir waves in plasma have been performed at the SB RAS Institute of Nuclear Physics (Novosibirsk) in the 1970s and 1980s (see, for example, Refs [9 - 12]). Of interest to us here is the formulation of the problem close to the classical Landau formulation [1, 2]. The problem involves a one-dimensional plasma system subject to periodic boundary conditions. The velocity distribution of plasma electrons is taken to be Maxwellian, and ions are assumed to be at rest ($m_i / m_e = 10^4$) and distributed uniformly over the length of the system. It is also assumed that at some initial point in time the system is subjected to small electron velocity and electron density perturbations of the form

$$\frac{\delta n}{n_0} = \frac{k_0 E_0}{4\pi e n_0} sin(\omega_0 t - k_0 x), \quad \delta v = \frac{\omega_0 E_0}{4\pi e n_0} sin(\omega_0 t - k_0 x), \tag{4.30}$$

corresponding to the linear wave $E(x,t) = E_0 \, sin(\omega_0 t - k_0 x)$, where $\omega_0^2 = \omega_e^2 + \frac{3}{2} k_0^2 v_T^2$, $k_0 = 2\pi / \lambda_0$, $v_T$ is the thermal electron velocity. The quantities $E_0$, $\varphi_0$, $\lambda_0$, $\omega_0$, and $k_0$ are the initial values of the field amplitude, potential, wavelength, frequency, and wave number, respectively. The numerical integration is performed using the `particles-in-cells' method. The number of particles is not large (in Refs [9 - 12], the authors usually put $N = 10^4$, with about $10^2$ particles per cell). To reduce the initial noise level, the `easy start' method is used, in which neither the coordinate nor velocity distribution functions of the electrons change from one cell to another. In this case, it was noted in Ref. [9] that the noise level is determined by computation errors but increases for the computation scheme chosen; the noise level increases with increasing $E_0$ and with decreasing $\lambda_0$. The computation only makes sense until the noise level remains small compared to the harmonics of the effect under study that arise in the calculation.

The calculations in Refs [9 - 12] were performed over a wide range of initial wave parameters. The time dependence of the field strength is quite complicated, but the initial stage always corresponds to the wave damping regime in which an increase in the amplitude and the phase velocity $v_\phi$ in the range $e\varphi_0 /(k_B T) > 1$ (and the corresponding decrease of the parameter $k_0 r_D$) dramatically increases the damping decrement compared to the Landau value (3.16).

In theory and mathematical experiments the question can be put about the number of trapped electrons by traveling wave electric field. But the mathematical expectation of the number of electrons moving with the phase velocity is zero. Then it is possible to speak only about a definite velocity interval, which contains phase velocity of wave, and about electrons (or probe particles in mathematical experiments) belonging to this interval.

In mentioned mathematical experiments they use also another possibility of calculation of the trapped particles. The mathematical experiments were realized with the initial number of particles in cell, which was not more than 100 and maximum initial velocity, which was less than $2{,}15\ v_T$, where $v_T = \sqrt{k_B T/m_e}$ is the thermal electron velocity. In experiments the phase velocity was changed in diapason from $2{,}46\ v_T$ to $22{,}4\ v_T$. Then it is possible to calculate the number of trapped probe particles which velocity – in the process of evolution of the physical system – become more than phase velocity $v_\phi$. At the beginning time moment all high energetic electrons in the tail of the Maxwell distribution function are cutting off. But the number of these "lost" electrons (estimation of authors of experiments) is not more than ~ 1,6%.

From discussed numerical experiments follow that the number of trapped electrons is not more than 20%. On the macroscopic level of the system description we can speak only about average (hydrodynamic) velocity $\bar{u}$ of electrons and about difference between $\bar{u}$ and $v_\phi$.

Fig. 33 summarizes the results of the numerical simulation in series experiments (3-1,…,3-6), [12]. Table 1 explains the used nomenclature.

**Table 1**. Nomenclature and data of the mathematical experiments by $\sqrt{e\langle\varphi\rangle/m_e} = const$

| Nomenclature Fig. 33 | triangles | oblique cross | black points | light points | straight cross | Rhombs |
|---|---|---|---|---|---|---|
| series | 3-1 | 3-2 | 3-3 | 3-4 | 3-5 | 3-6 |
| $\sqrt{e\langle\varphi\rangle/m_e}\,/v_T$ | 1 | 1.6 | 2.6 | 4.2 | 5.4 | 6.3 |
| $r_D k$ | 0.57-0.35 | 0.42-0.26 | 0.3-0.19 | 0.17-0.14 | 0.15-0.11 | 0.12-0.094 |

Black squares in Fig. 33 correspond to the data of the proposed analytical theory. From formulated numerical results of the mathematical modeling follow:

1. Reasonable coincidence of the damping decrement $\gamma$ with the Landau decrement $\gamma_L$ in series 3-1,…,3-6 is observed when potential energy of the electric field is of order of the heat energy and by rather large (but not small) values $r_D k$.

2. The results of the $\gamma/\gamma_L$ calculations depends only slightly on the ratio $\sqrt{e\langle\varphi\rangle/m_e}\,/v_T$ (in [12] $v_T = \sqrt{k_B T / m_e}$), but the difference between $\gamma$ and $\gamma_L$ catastrophically increases in spite of decreasing of the parameter $r_D k$. It is important that in series of the mathematical experiments (1-1,…,1-8) authors [12] come to a halt by the ratio $\gamma/\gamma_L$ ~$10^{104}$ by $r_D k$=0.045.

3. The results of the proposed analytical theory are in good coincidence with the data of the direct mathematical modeling. By the way the linear theory does not contain the explicit dependence on the electrical field energy.

This disagreement is easy to explain from the computational point of view. Let us consider the Landau formula (3.16) For $k_0 r_D$ <<1, the damping decrement calculated by Eqn (3.16) becomes very small, whereas its simulation counterpart does not differ much from the plasma frequency. Application of developed theory and relations (4.23) to the solution of classical problem of Landau damping makes it possible, even in Landau's linear formulation, to obtain a quite satisfactory agreement with mathematical experiments.

## Conclusion

As it was shown the Landau rule for analytical estimation of the corresponding singular integral should be considered as additional condition imposed on the physical system. Another condition with the transparent physical sense (connected with calculation of independent oscillations in physical system) leads to appearance of spectrum of oscillations.

The results of calculations are in good agreement with results of direct mathematical experiment even in Landau’s linear formulation, when the Landau formulae for decrement calculation leads to strong disagreement with results of direct mathematical experiments.

In spite of linear formulation of the problem the discussed complicated topology of curves (including the spectrum of the bell-like dispersion curves in Fig. 30 and Fig. 31) permits to give some conclusions about possible scenario of non-linear stage of process. The discrete spectrum of $\frac{dv_\phi}{d\lambda}$ and the discrete spectrum of $\frac{d(r_D k)}{d\hat{\omega}''}$ can lead to series of many successive components, which have some features of solitary waves. Moreover the Landau formulation contains two restrictions of principal significance – potential of the force field and maxwellian function as approximation for distribution function. Then the modified Landau formulation can be imposed in consideration of another physical system. From this point of view no surprise that appearance of successive solitary waves with velocities decreasing by about a factor of 2 was discovered in experiments in the principal different physical systems (see for example [14, 15]). But of course these effects need special thorough theoretical investigations.